\documentclass[sigconf]{acmart}

\usepackage{amsmath, amssymb, amsfonts}
\AtBeginDocument{%
  }
\usepackage{pifont}
\usepackage[normalem]{ulem}
\usepackage{colortbl}
\usepackage{xcolor}
\usepackage{enumitem}
\usepackage{todonotes}
\usepackage{booktabs} 
\usepackage{multirow}
\usepackage{tabularx}
\usepackage{makecell}
\usepackage{tikz}
\usepackage{graphicx}
\usepackage{subcaption}
\usepackage{tikz}
\newcommand*\circled[1]{\tikz[baseline=(char.base)]{
            \node[shape=circle,draw,inner sep=0.4pt, minimum size=12.8pt] (char) {\vphantom{1g}\normalfont\small\textbf{#1}};}}
\copyrightyear{2025}
\acmYear{2025}
\setcopyright{acmlicensed}\acmConference[CHI '25]{CHI Conference on Human Factors in Computing Systems}{April 26-May 1, 2025}{Yokohama, Japan}
\acmBooktitle{CHI Conference on Human Factors in Computing Systems (CHI '25), April 26-May 1, 2025, Yokohama, Japan}
\acmDOI{10.1145/3706598.3713682}
\acmISBN{979-8-4007-1394-1/25/04}

\begin{document}

\title{CoKnowledge: Supporting Assimilation of Time-synced Collective Knowledge in Online Science Videos}



\author{Yuanhao Zhang}
\email{yzhangiy@connect.ust.hk}
\affiliation{%
  \institution{Hong Kong University of Science and Technology}
  \city{Hong Kong}
  \country{China}
}

\author{Yumeng Wang}
\email{ywanglu@connect.ust.hk}
\affiliation{%
  \institution{Hong Kong University of Science and Technology}
  \city{Hong Kong}
  \country{China}
}

\author{Xiyuan Wang}
\email{wangxy7@shanghaitech.edu.cn}
\affiliation{%
  \institution{ShanghaiTech University}
  \city{Shanghai}
  \country{China}
}

\author{Changyang He}
\email{changyang.he@mpi-sp.org}
\affiliation{%
  \institution{Max Planck Institute for Security and Privacy}
  \city{Bochum}
  \country{Germany}
}

\author{Chenliang Huang}
\email{clhuang77@gmail.com}
\affiliation{%
  \institution{New York University}
  \city{New York}
  \country{United States}
}

\author{Xiaojuan Ma}
\email{mxj@cse.ust.hk}
\affiliation{%
  \institution{Hong Kong University of Science and Technology}
  \city{Hong Kong}
  \country{China}
}
\renewcommand{\shortauthors}{Zhang et al.}

\begin{abstract}
  % In online video-sharing platforms, viewers engage with science content through danmaku, a system of scene-aligned, time-synced comments that, together with the video content, constitute “collective knowledge.” However, the inherent nature of danmaku often impedes viewers' ability to effectively assimilate this collective knowledge in science videos. With a formative study, we examined viewers' practices for processing collective knowledge and the specific barriers they encountered. Building on these insights, we designed a processing pipeline to filter, classify, and cluster danmaku, leading to the development of CoKnowledge—a tool incorporating video abstracts, knowledge graphs, and supplementary configurations. Through a within-subject study (N=24), CoKnowledge could significantly enhance participants’ comprehension and recall of collective knowledge compared to a Bilibili-like baseline. Additionally, we analyzed user interaction patterns and perceptions of the design elements. Finally, we present design considerations for developing similar support tools.
  Danmaku, a system of scene-aligned, time-synced, floating comments, can augment video content to create `collective knowledge'. However, its chaotic nature often hinders viewers from effectively assimilating the collective knowledge, especially in knowledge-intensive science videos. With a formative study, we examined viewers' practices for processing collective knowledge and the specific barriers they encountered. Building on these insights, we designed a processing pipeline to filter, classify, and cluster danmaku, leading to the development of CoKnowledge -- a tool incorporating a video abstract, knowledge graphs, and supplementary danmaku features to support viewers' assimilation of collective knowledge in science videos. A within-subject study (N=24) showed that CoKnowledge significantly enhanced participants’ comprehension and recall of collective knowledge compared to a baseline with unprocessed live comments. Based on our analysis of user interaction patterns and feedback on design features, we presented design considerations for developing similar support tools. 

\end{abstract}

\begin{CCSXML}
<ccs2012>
   <concept>
       <concept_id>10003120.10003121.10003129</concept_id>
       <concept_desc>Human-centered computing~Interactive systems and tools</concept_desc>
       <concept_significance>500</concept_significance>
       </concept>
   <concept>
       <concept_id>10003120.10003121.10011748</concept_id>
       <concept_desc>Human-centered computing~Empirical studies in HCI</concept_desc>
       <concept_significance>300</concept_significance>
       </concept>
 </ccs2012>
\end{CCSXML}

\ccsdesc[500]{Human-centered computing~Interactive systems and tools}
\ccsdesc[300]{Human-centered computing~Empirical studies in HCI}

\keywords{Collective Knowledge, Online Video Platforms, Danmaku, Science Communication}


\begin{teaserfigure}
  \includegraphics[width=\textwidth]{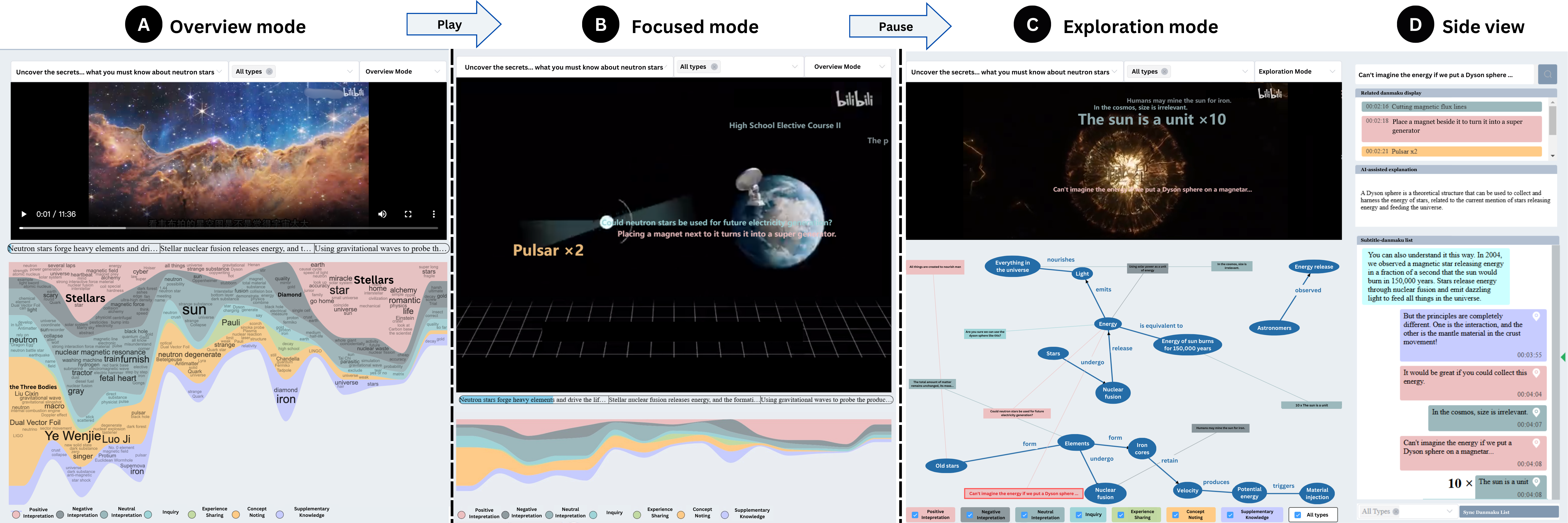}
  \caption{Overview of design features in CoKnowledge: A) Overview mode integrated a progress bar directory and Wordstream for video knowledge abstraction. B) Focused mode enlarged the video window with minimal interactive features. C) Exploration mode provided timestamped knowledge graphs for structured analysis. D) Side view, constant across all modes, offered supplementary danmaku features, including related danmaku display, AI-assisted explanations, and a subtitle-danmaku list.}
  \Description{Overview of design features in CoKnowledge: A) Overview mode integrated a progress bar directory and Wordstream for video knowledge abstraction. B) Focused mode enlarged the video window with minimal interactive features. C) Exploration mode provided timestamped knowledge graphs for structured analysis. D) Side view, constant across all modes, offered supplementary danmaku features, including related danmaku display, AI-assisted explanations, and a subtitle-danmaku list. The system defaulted to Overview mode when users just loaded a video. When users began playing the video, CoKnowledge automatically transitioned to Focused mode. When users paused at a specific timestamp, CoKnowledge entered Exploration mode.}
  \label{fig:teaser}
\end{teaserfigure}

% \begin{figure}[h]
%   \centering
%   \includegraphics[width=\linewidth]{images/teaser.png}
%   \caption{User interface of Coknowledge...\xm{the subfigures in each mode are too small to read...}}\label{fig:teaser}
% \end{figure}

\maketitle

\section{Introduction}

\begin{figure*}[h]
  \centering
  \includegraphics[width=0.7\linewidth]{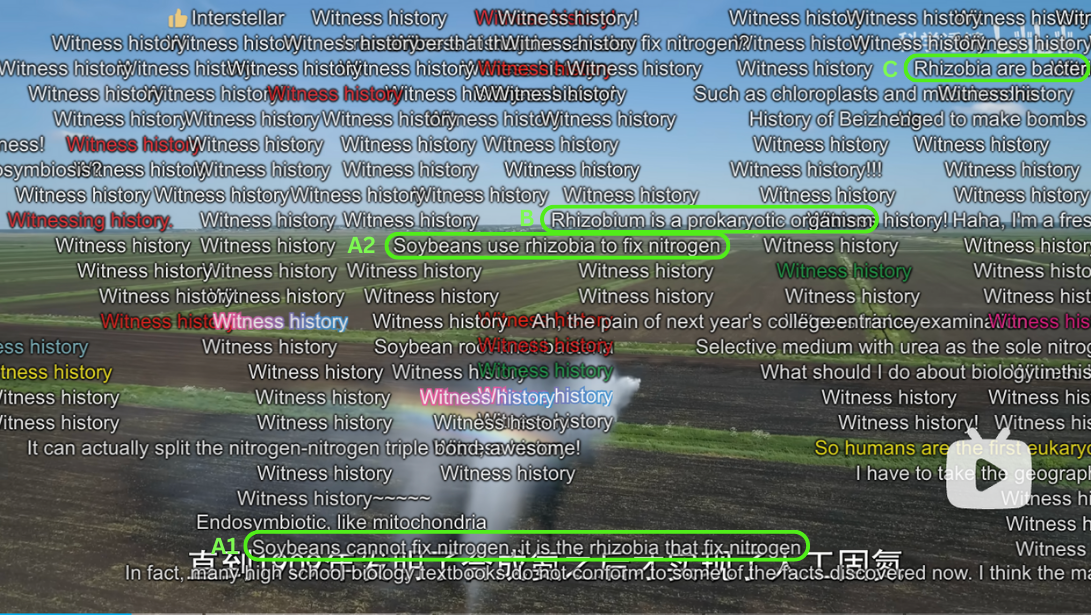}
  \caption{The screenshot shows scrolling danmaku comments overlaid on a video discussing the discovery of the first eukaryotic organism capable of nitrogen fixation. Comments A1 and A2 state the same thing: that rhizobia, not soybeans, fix nitrogen. Comments B and C further explain that rhizobia are bacteria/prokaryotes. These complementary danmaku enhance the breadth and depth of the video knowledge. However, the large volume of danmaku, most of which lack informational content, distracts viewers from the video content and makes it difficult to notice the \textit{knowledge danmaku}.}
  \Description{The screenshot shows scrolling danmaku comments overlaid on a video discussing the discovery of the first eukaryotic organism capable of nitrogen fixation. Comments A1 and A2 state the same thing: that rhizobia, not soybeans, fix nitrogen. Comments B and C further explain that rhizobia are bacteria/prokaryotes. These complementary danmaku enhance the breadth and depth of the video knowledge. However, the large volume of danmaku, most of which lack informational content, distracts viewers from the video content, and makes it difficult to notice the knowledge danmaku.}
  \label{fig:danmaku}
\end{figure*}

Online video-sharing platforms, with billions of active users daily \cite{cisco2018cisco}, are playing an increasingly important role in engaging the general public in science communication \cite{morcillo2015typologies,welbourne2016science,xia2022millions}. 
These platforms enable science communicators to capture public attention through cinematic techniques and rich, dynamic, multimodal narratives \cite{zhang2023understanding,huang2020good,morcillo2015typologies}. They encourage viewers to share their thoughts on scientific topics and interact with both the content creator and other audiences through commenting features \cite{zhang2023understanding}.
In particular, danmaku, a distinctive
% novel \xm{emerging? it has been around for a while...} 
commentary system widely adopted in online video platforms like Bilibili \cite{bilibili}, Niconico \cite{niconico}, and Douyin \cite{douyin}
% for online videos \xm{widely used in what platforms?}
, allows anonymous posting \cite{wu2019danmaku} of scene-aligned comments for a fixed display period on the video timeline regardless of the actual post time \cite{ma2017video}. 
% \mm{These platforms enable science communicators to capture public attention through cinematic techniques and rich, dynamic, multimodal narratives. They encourage viewers to share their thoughts on scientific topics and interact with both content creators and other audiences through commenting features, fostering a sense of `collective knowledge'. A notable example is danmaku, a distinctive commentary system widely adopted in online video platforms like Bilibili \cite{bilibili}, Niconico \cite{niconico}, Douyin \cite{douyin}. This system allows anonymous posting of scene-aligned comments for a fixed display period on the video timeline, regardless of the actual post time.}
These comments scroll from right to left across the viewing window (Figure \ref{fig:danmaku}) 
\cite{yao2017understanding,he2021beyond,wu2019danmaku,ma2017video}, creating a sense of synchronous, crowd-generated 
% \xm{correct?} 
textual augmentation to video content, directly influencing how viewers perceive and digest the videos  \cite{he2021beyond}.

Although most prior research on danmaku has focused on its role in engaging and entertaining viewers by reducing the feeling of isolation \cite{liao2023research,chen2015understanding,lin2018exploratory}, it has also proven to be an effective channel for knowledge sharing and information seeking \cite{he2021beyond, li2022classification, ma2017video, huang2020good, chen2017watching}. 
Danmaku fosters a sense of watching or even participating in live discussions around the corresponding video scenes \cite{chen2015understanding,ma2014analysis, huang2024sharing}. 
Compared to traditional comments, which typically offer general and after-the-fact insights, danmaku delivers a comparable volume of knowledge that enriches and extends the video content with greater detail, specificity, and real-time relevance \cite{he2021beyond, wu2018danmaku, wu2019danmaku, chen2017watching}.
Such floating comments can be viewed as a form of knowledge co-creation. This paper defines the aggregation of
% \xm{topic-relevant?} 
topic-relevant information co-contributed by video content and temporally anchored commentary augmentations as \textbf{`time-synced collective knowledge'} (referred to as `collective knowledge' for short). With tight correspondence between danmaku and specific video content \cite{he2021beyond}, such collective knowledge ensures that the information presented in science videos is not solely reliant on the creator, thereby reducing potential biases \cite{he2021beyond,yao2017understanding}.
% \xm{Highlight why collective knowledge is particular valuable for science videos with references, if possible.}
%We define this robust and comprehensive co-constructed knowledge structure as 'time-synced collective knowledge,' which is the aggregation of information co-created by both video content and time-synced augmentations via danmaku. ``''
% \mm{While much of the prior research on danmaku has focused on its role in engaging and entertaining viewers by reducing feelings of isolation, it has also proven to be an effective channel for knowledge sharing and information seeking. Danmaku fosters a sense of participation, allowing viewers to engage in live discussions around the corresponding video scenes. Among the floating comments, those aiming to enrich and extend the video content can be viewed as a form of knowledge co-creation. In this paper, we define the aggregation of topic-relevant information co-contributed by video content and temporally anchored commentary augmentations as “time-synced collective knowledge.” Such collective knowledge ensures that the information presented in science videos is not solely reliant on the creator, thereby reducing potential biases.}

However, viewers of science videos often find it challenging to assimilate such collective knowledge due to the nature of danmaku \cite{chen2015understanding, chen2024towards,ma2017video,yao2017understanding}. The abundance and uneven quality of live comments can interfere with and detract people from the original video content \cite{chen2015understanding,chen2024towards}. Additionally, the high volume of danmaku means that individual pieces may not receive sufficient attention \cite{ma2017video,yao2017understanding}. Therefore, many viewers tend to ignore or turn off danmaku and thus miss out on valuable knowledge scattered around the comments.
%Despite the importance of collective knowledge, we have noticed limited Human-Computer Interaction (HCI) works to support users' assimilation of collective knowledge in online videos. There are existing studies that examine the processing of danmaku in online videos. One example is VideoForest \cite{sun2016videoforest}, which utilizes danmaku data to summarize video streams, aiding in the identification of keyframes associated with danmaku comments. However, in this context, danmaku is considered secondary to the video data and is not recognized as a substantial source of knowledge. Consequently, there is a lack of in-depth analyses of danmaku knowledge. On the other hand, DanmuVis enables the analysis of danmaku content and viewers' behaviors in relation to both video time and post time \cite{chen2022danmuvis}, providing insights into viewers' online participation and perceived experience. Nevertheless, its design primarily caters to video uploaders seeking to understand the dynamics of their videos, rather than supporting viewers in assimilating the knowledge encompassed within the danmaku and video content. 
Some existing studies examined the use of danmaku data to facilitate video content processing, such as identifying key scenes in videos based on danmaku volume and sentiment \cite{sun2016videoforest}, and analyzing viewer behavior \cite{chen2022danmuvis}. 
Nevertheless, prior work primarily catered to understanding video dynamics and user engagement rather than supporting viewers in assimilating the knowledge encompassed within danmaku comments and video narratives. 
The intricate interplay between danmaku and science video content, as well as their underlying knowledge patterns, remains largely unexplored.

To address this gap, we first conducted a \hyperref[formative-study]{Formative Study} with ten frequent viewers of online science videos that feature danmaku
% \xm{with whom} 
to identify users' needs and challenges for absorbing and integrating collective knowledge while watching science videos.
%comprehensively understand how users assimilate collective knowledge in online science videos. 
Our results confirmed that danmaku is a viable knowledge source. 
By analyzing participants' reflections on their current practices, 
% \xm{correct?}
we pinpointed two main obstacles to benefiting from collective knowledge embedded in danmaku: low information density and an obscure information structure. 
%Subsequently, we identified users' behavior patterns as they assimilate collective knowledge, as well as the obstacles encountered during this process. Two key challenges associated with danmaku emerged: low information density and an obscure information structure. These findings prompted us to conduct a qualitative analysis with a danmaku pool (aligned with the corresponding video content) to explore the knowledge patterns of danmaku. 
These findings prompted us to conduct a qualitative analysis of a collection of science videos and their danmaku comments to explore the patterns of viewer-contributed knowledge.
%We proposed the concept of \textit{'knowledge danmaku'}, which refers to danmaku that facilitates the construction or assimilation of collective knowledge. We also identified their information themes relative to video content. 
We distilled \textit{`knowledge danmaku'} -- live comments that facilitate the construction or digestion of collective knowledge -- and derived their information themes with regard to the associated video content using \hyperref[content-analysis]{Content Analysis}.
% \xm{what qualitative research method?}
Using these labeled data for fine-tuning,
% \xm{correct?}`'
we established a natural language processing (NLP) pipeline that automatically extracted \textit{knowledge danmaku} of a science video, classified them according to their information categories, 
and finally clustered danmaku within the same category by semantic similarity.
% \yh{and finally mapped the comments to their corresponding video knowledge segment.}
With this pipeline as the backbone and guided by the insights from the formative study, we further designed and implemented CoKnowledge, an interactive system to support viewers' assimilation of collective knowledge in online science videos.
CoKnowledge proposed to achieve this goal by 1) providing a high-level knowledge abstraction for efficient skimming and locating, 2) constructing knowledge graphs at each video timestamp for structured analysis, and 3) presenting supplementary features of each danmaku for exploration and meaning-making.
% thorough exploration.
%Building on the insights from the formative study and qualitative analysis, we further designed and implemented CoKnowledge, an interactive system to support viewers' assimilation of collective knowledge in online science videos. CoKnowledge is founded on a Natural language processing (NLP) pipeline that first filtered knowledge danmaku of a science video from all its danmaku based on the definition, then classified them according to their information themes, and finally clustered danmaku of similar semantics  within the same theme. Based on the pipeline, CoKnowledge fostered viewers' assimilation to collective knowledge by 1) providing a high-level video abstract for efficient skimming and navigation, 2) constructing knowledge graphs at each video timestamp for structured analysis, and 3) presenting supplementary features of each danmaku for thorough exploration.

We conducted a within-subject study with 24 participants to evaluate CoKnowledge against a baseline system that emulates the interface of Bilibili (i.e., a popular video-sharing platform with danmaku features in China \cite{bilibili_about_us}). Our evaluation focused on three key aspects: 1) CoKnowledge's efficacy in supporting viewers' assimilation of collective knowledge and maximizing danmaku's capabilities, 2) users' interaction with CoKnowledge and their perceptions of its design features, and 3) CoKnowledge's overall usability. The results of mixed-methods analyses -- including in-task quizzes, in-task surveys, and interviews -- demonstrated that CoKnowledge significantly enhanced users' comprehension and recall of the collective knowledge. It was also perceived as more effective in leveraging the knowledge co-construction capabilities of danmaku 
% \xm{correct?}
than the baseline. %It was also perceived as more effective in supporting knowledge assimilation and leveraging the capabilities of danmaku. 
We identified behavior patterns that emerged during users' interaction with CoKnowledge, and participants reported high perceived usefulness for most of the design elements. Moreover, despite the additional interactive features, CoKnowledge still maintained comparable usability and task workload levels to the baseline. Based on our findings, we further summarized practical design implications for future tools similar to CoKnowledge.

The key contributions of this work include
% \xm{may convert to regular paragraph to save space}: 
\textbf{1)} CoKnowledge, an interactive system that supported viewers' assimilation of collective knowledge in online science videos; \textbf{2)} a mixed-methods user study investigating the usefulness and usability of CoKnowledge while exploring viewers' interaction patterns; and \textbf{3)} design considerations to guide future support for assimilating knowledge co-constructed by media producers and consumers.

\section{Related Work}

\subsection{Knowledge Co-Construction in Online Science Videos}
Scientific understanding plays a crucial role in our daily lives, prompting many individuals to seek information online \cite{tabak2015functional, segev2012seeking, brossard2013new, dubovi2020empirical}. With the rise of video-sharing platforms, videos have emerged as a widely used medium for science communication \cite{morcillo2015typologies, welbourne2016science}. Unlike traditional texts or static images, videos offer a dynamic way to present scientific knowledge, utilizing visual aids and narrative techniques to simplify complex concepts for general audiences while maintaining scientific integrity \cite{trumbo1999visual, nicholson2005representing}. This approach not only captures public attention but also enhances the accessibility of science \cite{zhang2023understanding}.

A notable feature of video-sharing platforms is the interactive environment they foster, allowing viewers to post comments on video content. The commentary feature extends beyond passive consumption, enabling viewers to participate actively in science communication by sharing their insights and engaging in discussions with both the creator and other viewers \cite{he2021beyond, wu2018danmaku, lange2007publicly, dubovi2020empirical}. Such information exchange often leads to knowledge co-construction, where individuals not only share perspectives but also collaboratively refine and expand on each other's ideas, thereby enhancing the collective understanding \cite{he2021beyond, dubovi2020empirical, scardamalia2002collective}. This co-constructed knowledge becomes a valuable resource for both viewers, who gain additional video-related information \cite{choi2020finding}, and creators, who receive timely feedback \cite{wu2018danmaku}. The importance of such co-constructed knowledge is particularly pronounced in science videos, where the accuracy of information is paramount \cite{stadtler2007dealing, dubovi2020empirical}, and the risk of misinformation is significant \cite{rosenthal2020media}. Therefore, our work focused on collective knowledge arising from online science videos.

% \xm{shall we mention CK for genres beyond science videos, e.g., how-to videos?}

\subsection{Danmaku for Knowledge Co-Construction}
Danmaku, a commentary system originating from Japan, has gained widespread popularity on Asian video platforms, particularly on Bilibili, a popular danmaku platform in China \cite{bilibili_about_us}. By the first quarter of 2024, Bilibili had amassed 341 million monthly active users \cite{bilibili2024investor} and ranked third globally among the most visited Arts and Entertainment websites \cite{similarweb_top}, behind YouTube \cite{youtube} and Netflix \cite{netflix}. 
In June 2020, Bilibili introduced the "Knowledge Zone" channel to curate science knowledge content. The channel experienced a 92\% increase in creators within a year \cite{bilibili2021report}. By 2024, it has accumulated over 8.5 million science-related videos \cite{bilibili2023esg}.

Unlike traditional comments displayed asynchronously below the video, danmaku comments are displayed directly on the video, scrolling horizontally across the screen. These comments are synchronized with specific video scenes, creating a co-viewing experience where users can send and view comments anonymously and simultaneously while watching the video \cite{chen2017watching, chen2024towards}. 
% \xm{[Again, highlight here that how danmaku contributes new knowledge beyond video captions, subtitles, and traditional comments (R1)]}

Research suggests that danmaku serves as a powerful medium for knowledge dissemination \cite{ma2014analysis}. 
Wu et al. highlighted that compared to traditional forum-style comments, danmaku excels in disseminating explicit knowledge \cite{wu2018danmaku, wu2019danmaku}.
Moreover, while traditional comments tend to initiate general discussions of overarching video topics, danmaku addresses specific video details, offering timely supplementation and corrections that facilitate more fine-grained knowledge co-construction \cite{he2021beyond, wu2018danmaku, chen2017watching}.
On the one hand, danmaku can respond to specific video scenes by providing immediate explanations (e.g., defining terms \cite{li2015can} or crowdsourcing captions \cite{ma2014analysis}), thus expanding the scope of knowledge presented. On the other hand, danmaku allows users to engage in discussions that correct or enhance previous comments, which deepens the knowledge shared \cite{he2021beyond}. Even in cases of disagreement, the range of perspectives available through danmaku contributes to a richer understanding of the content \cite{wu2018danmaku, wu2019danmaku}. Such a dynamic, collective knowledge system is often more comprehensive and resilient than traditional videos, where the creator is the sole source of information \cite{yao2017understanding}.
However, collective knowledge patterns in science videos, a knowledge-intensive genre, remain underexplored.
% Taking inspiration from these previous works, our study is located within the context of science videos, a knowledge-intensive genre. 
Through content analysis, we investigated how danmaku contributed to knowledge co-creation with science videos and explored its potential as a valuable source of collective knowledge in science communication.

\subsection{Technologies for Processing Danmaku and Facilitating UGC Assimilation}
Danmaku has garnered increasing attention from video researchers in the HCI and CSCW communities \cite{ma2017video, wu2018danmaku, he2021beyond, wu2019danmaku}. Several tools have been developed to leverage the pseudo-synchronous nature of danmaku for various purposes. Sun \textit{et al.} introduced VideoForest \cite{sun2016videoforest}, a system that utilizes danmaku data to summarize video streams and identify key frames aligned with user comments. Similarly, Cao \textit{et al.} developed VisDmk \cite{cao2023visdmk}, an interactive visual analysis tool designed to explore emotional trends and characteristics of danmaku comments as a video unfolds. This system helps users gain an overview of the video, evaluate its impact, and expedite the video editing process. To assist video creators in understanding viewer engagement, Chen \textit{et al.} built DanmuVis \cite{chen2022danmuvis}, which analyzes danmaku content and viewer behaviors over time, providing insights into online participation and perceived user experience.
Although these studies offer insights into the dynamics of danmaku, their primary focus remains on enhancing video analysis. In contrast, our work shifted the emphasis toward the intrinsic knowledge value of danmaku itself, developing an NLP pipeline and an interactive interface to maximize its potential for knowledge co-creation.

% In general, online User-generated content (UGC) serves as a valuable resource for individuals seeking to learn from others, which makes the effective digestion of UGC a prominent research topic. While there is limited research on how to help viewers effectively assimilate collective knowledge from online science videos, insights can be drawn from existing works on UGC digestion, particularly given that danmaku is a form of UGC. 
While there is limited research to support the assimilation of collective knowledge in science videos, insights can be drawn from existing works on user-generated content (UGC) digestion, particularly given that danmaku is a form of UGC. 
In general, UGC serves as a valuable resource for individuals seeking to learn from others, which makes the effective consumption of UGC a prominent research topic. 
One major approach to supporting UGC digestion involves providing text summarization \cite{dave2004flash, hoque2015convisit}. For instance, CoArgure offers a holistic view of collective arguments \cite{liu2023coargue} through claim centers, facilitating the processing of content from community question-answering (CQA) platforms. Due to the unstructured nature of UGC, presenting it in a more organized manner can reduce cognitive load and aid comprehension \cite{liu2022planhelper}. Kitayama \textit{et al.} proposed a video-sharing system that organizes comments by their corresponding video time intervals and screen areas \cite{kitayama2008organizing}, which improves users’ understanding of video comments. Similarly, Liu \textit{et al.} designed PlanHelper to organize CQA answers \cite{liu2022planhelper} into a list and a mindmap, helping users construct activity plans more efficiently.
% \xm{unique challenges of danmaku compared to other kinds of UGC?}
However, compared to other forms of UGC, danmaku poses unique challenges due to its uneven quality, transient display, and random appearance across the screen, making it harder for users to absorb the embedded knowledge \cite{ma2017video, yao2017understanding}.
% To further investigate these challenges, we conducted a formative study with consumers of online science videos featuring danmaku. Based on our findings and insights from previous works, we derived the design pipeline for developing CoKnowledge.

\section{Formative Study}
\label{formative-study}
% To understand viewer behavior and challenges in assimilating collective knowledge in science videos, we conducted semi-structured interviews with 10 frequent consumers of these videos with danmaku feature. Our study validates the benefits of danmaku in science videos, categorizing three user behavior modes and identifying the challenges encountered. 
% % These insights informed the design of \textit{SystemName}, an interactive system to aid viewer comprehension of collective knowledge.
To further investigate the needs and challenges specific to danmaku, we conducted a formative study with consumers of science videos featuring danmaku. Based on our findings and insights from previous works, we derived the design pipeline for CoKnowledge.

\subsection{Participants and Procedure}
With IRB approval, we recruited 10 participants (five female, five male; labeled PF1 to PF10) for the formative study through online ads, social media, and word-of-mouth. All participants were self-identified frequent viewers of online science videos with the danmaku feature (six watched daily, three watched 4 to 6 days a week, and one watched at least once a week).  In semi-structured interviews, participants discussed their viewing strategies, interactions with danmaku, and perceptions of its role in collective knowledge. We also explored challenges in assimilating collective knowledge and solicited suggestions for technical support. Participants were encouraged to provide examples to contextualize their responses. 
% Interviews lasted 30-45 minutes, and participants received \$6 for their time.
Following established protocols \cite{klein2017quality, baker2021avatar, liu2021makes}, interviews were audio-recorded, transcribed, and thematically analyzed by two researchers.
% Two researchers conducted a thematic analysis to inductively code and identify themes. 
We monitored the codebook for saturation after eight interviews \cite{saunders2018saturation}, completing the last two to confirm that no new themes emerged.

\subsection{Summary of Findings}
Our findings validate danmaku's benefits in science videos, identifying three user behavior modes and challenges in consuming collective knowledge.

\begin{table*}[h!]
\centering
\scalebox{0.75}{ 
\begin{tabular}{ >{\centering\arraybackslash}m{1.2cm} >{\centering\arraybackslash}m{5cm} >{\centering\arraybackslash}m{4.8cm} >{\centering\arraybackslash}m{4cm} >{\centering\arraybackslash}m{3.5cm} }
  \hline
   &  & \textit{Overview} & \textit{Focused} & \textit{Exploration} \\
  \hline
  \multirow{5}{*}{Dynamics} & Knowledge need & High-level knowledge abstraction \circled{1}& Concise knowledge \circled{2}& Structured knowledge \circled{3}\\
  & Knowledge scope & Entire video \circled{4}& Current timestamp \circled{5}& Short segments \circled{6}\\
  & Engagement level & Active \circled{7}& Mostly passive \circled{8}& Active \circled{9}\\
  & Interaction behavior & Skimming, locating \circled{10}& Typical viewing behavior \circled{11}& Thorough analysis \circled{12}\\
  & Focal point & Entire interface \circled{13}& Video segment view \circled{14}& Entire interface \circled{15}\\
  \hline
  \multirow{5}{*}{Challenges} & Mutual distractions \circled{16}& \ding{55} & \ding{51} & \ding{55} \\
  & Short display duration \circled{17}& \ding{55} & \ding{51} & \ding{55} \\
  & Danmaku locating difficulties \circled{18}& \ding{51} & \ding{55} & \ding{55} \\
  & \textbf{Low information density} & \ding{51} & \ding{51} & \ding{51} \\
  & \textbf{Obscure information structure} & \ding{51} & \ding{51} & \ding{51} \\
  \hline
\end{tabular}
}
\caption{Behavior patterns and challenges for assimilating collective knowledge in online science videos. The labeled items are addressed in section \ref{ui}.}
\label{formative-study-findings}
\end{table*}

\subsubsection{Advantages of Danmaku}

We inquired about participants' perceptions of the support provided by danmaku in knowledge acquisition while watching science videos. 
% They highlighted numerous advantages of danmaku, including those mentioned in existing work. 
% Most participants (9 out of 10) indicated that danmaku offered a \textbf{sense of co-presence}. PF9 remarked, \textit{``When danmaku is turned off, the interaction feels like a conversation between just me and the uploader. When it’s on, everyone is involved in discussing the topic.''} 
The majority (9 out of 10) highlighted that danmaku fostered a \textbf{sense of co-presence}, with PF9 noting it shifted the experience from an ``isolated interaction with the uploader'' to a more collective discussion.
This co-presence enhanced engagement with the content, thereby boosting motivation and efficiency in knowledge acquisition.
% Furthermore, 8 participants emphasized that danmaku helped \textbf{avoid knowledge bias} by presenting diverse viewpoints. Specifically, PF8 mentioned, \textit{``Sometimes I worry that the uploader's perspective might be biased, but danmaku provides a variety of viewpoints.''} These diverse perspectives offered users additional references, facilitating a comprehensive understanding of the content.
Furthermore, 8 participants emphasized that danmaku helped \textbf{avoid knowledge bias} by presenting diverse viewpoints, as PF8 indicated, offering ``alternative perspectives'' to those of the uploader.
% All participants agreed that danmaku synchronously supplemented knowledge. PF2 stated, \textit{``[Compared to comments,] I can see real-time, immediate supplements or extensions of the video content through danmaku.''} 
All participants agreed that danmaku synchronously supplemented knowledge, with PF2 highlighting its ability to offer ``real-time'' and ``immediate'' extensions of video content.
\textbf{The closer integration} of danmaku with the video enhanced the immediacy and contextual relevance of the knowledge.
These advantages aligned with those discussed in existing literature \cite{he2021beyond, yao2017understanding, chen2015understanding, ma2014analysis, huang2024sharing}, further affirming danmaku as a valuable source for constructing collective knowledge.
% In addition to the advantages mentioned in previous work, participants also identified other benefits of danmaku. For instance, PF5 noted, \textit{``I can find a sense of agreement in the viewpoints expressed in danmaku,''} while PF7 mentioned that the flow of danmaku helps them stay focused and reduces distractions. These advantages confirm danmaku as a valuable source for constructing collective knowledge.

\subsubsection{Viewing Practices and Challenges}

Participants reported various practices when assimilating collective knowledge. We identified three viewing modes—\textbf{Overview}, \textbf{Focused}, and \textbf{Exploration}—and examined their corresponding features (Table \ref{formative-study-findings}).

In the \textbf{Overview} mode, many participants (PF1, PF4-7, PF10) actively skimmed through the video to quickly understand or review its main content and locate segments of interest. Such behaviors were often observed before watching the video, but also during (PF6) and after viewing (PF6, PF7, PF10). However, users struggled with effectively \textbf{locating desired danmaku content}. PF1 remarked, \textit{``There is no way to preview danmaku to locate the good ones,''} PF6 added, \textit{``I saw interesting danmaku, but when I wanted to revisit them later, I could not find them.''}

The \textbf{Focused} mode described the situation when users concentrated solely on watching the video. In this mode, users were mostly passive, occasionally engaging in typical viewing behaviors like pausing, replaying, skipping, or adjusting playback speed to shift into other modes. All participants reported being overwhelmed by the volume of danmaku and finding the danmaku and video to be \textbf{mutually distracting}. Danmaku often blocked the video content (PF1, PF2-5, PF8), causing users’ attention to oscillate between the two, significantly hindering the assimilation of collective knowledge. Many participants (7 out of 10) also complained about the \textbf{short display duration} of danmaku, with PF10 noting that comments \textit{``fly by too quickly''} to be fully read.

When participants encountered interesting or complex content in the video or danmaku, they entered the \textbf{Exploration} mode. In this mode, users fixated on a short segment of the video, examining the meaning of danmaku and its relationship with the video (PF6, PF10), and may also seek additional information from external sources to aid their understanding (PF1, PF2, PF10).

We identified two key challenges that fundamentally hinder the assimilation of collective knowledge across all viewing modes: \textbf{low information density} and \textbf{obscure information structure}, which were frequently criticized by participants. All participants agreed that most danmaku is \textit{``meaningless for knowledge acquisition''} and characterized by excessive repetition, often burying valuable comments (PF1-6, PF7, PF9) and causing frustration (PF3). Furthermore, the majority (8 out of 10) mentioned that even informative danmaku was chaotic because the \textit{``information pattern was unorganized''} (PF7), and the relationship between danmaku and video knowledge lacked clarity (PF10), making it challenging to synthesize danmaku into the framework of collective knowledge.

\subsection{Design Requirements}
\label{drs}

\begin{figure*}[h]
  \centering
  \includegraphics[width=\linewidth]{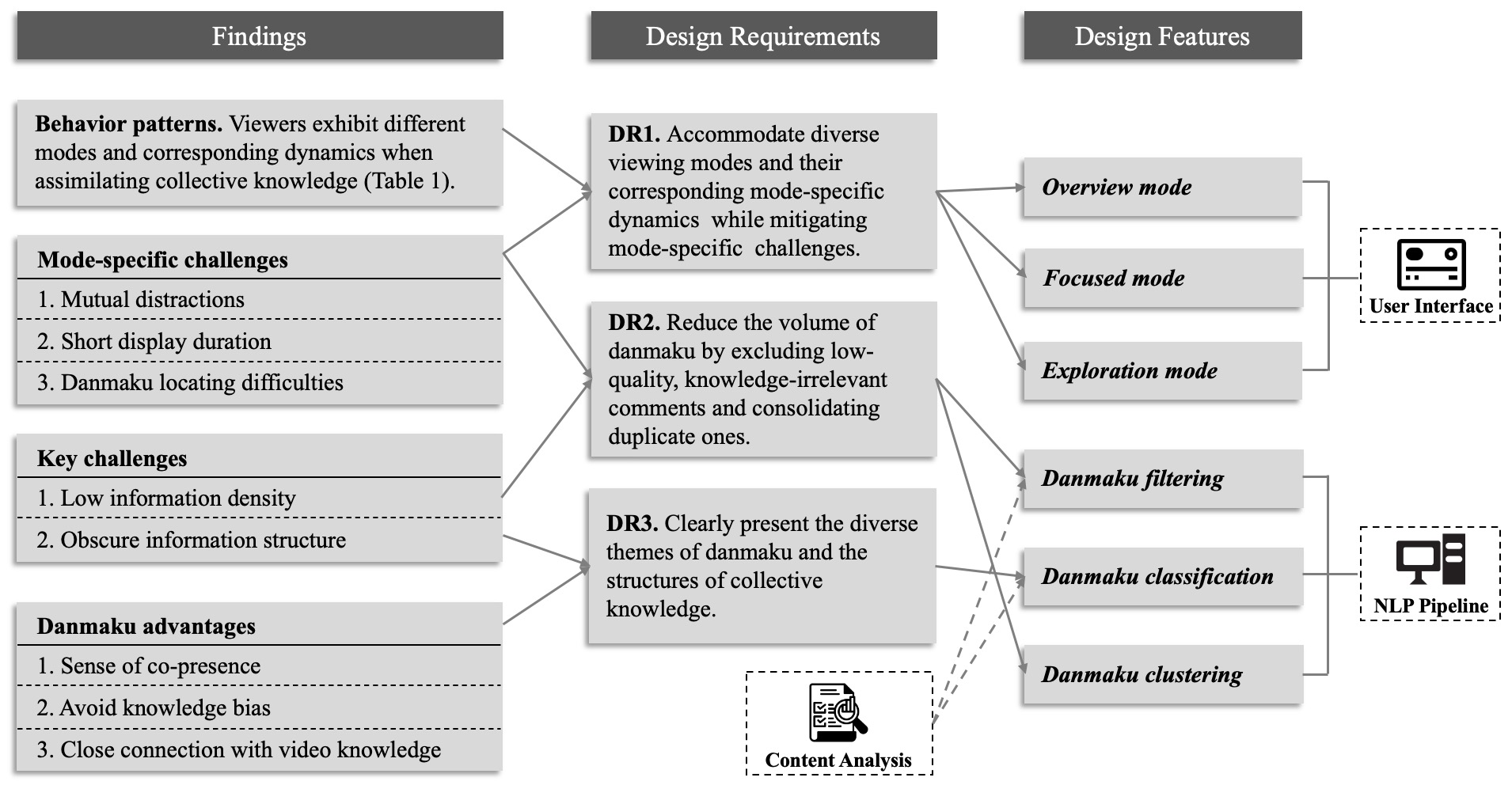}
  \caption{This figure illustrates an overview of our design pipeline: based on the findings derived from the formative study, we established design requirements to address each finding. Subsequently, we developed corresponding features that constituted the components of our design to meet these requirements.}\label{fig:design-pipeline}
  \Description{This figure illustrates an overview of our design pipeline: based on the findings derived from the formative study, we established design requirements to address each finding. Subsequently, we developed corresponding features that constituted the components of our design to meet these requirements. Specifically, let's start with the correspondence of findings to design requirements. Various behavior patterns and mode-specific challenges (Mutual distractions, short display duration, and danmaku locating difficulties) correspond to DR1. Mode-specific challenges and low information density correspond to DR2. Obscure information structure and danmaku advantages correspond to DR3. Then let's go to the correspondence of design requirements to design features. DR1 corresponds to Overview, Focused, and Exploration modes. DR2 corresponds to danmaku filtering and clustering. DR3 corresponds to danmaku classification. And we also conducted a content analysis to assist automatic danmaku filtering and classification. The Overview, Focused, and Exploration modes constitue the user interface of CoKnowledge. The danmaku filtering, classification and clustering constitue the NLP pipeline of CoKnowledge.}
\end{figure*}

Based on the findings, we formulated the following design requirements to guide the subsequent design process (as illustrated in Fig \ref{fig:design-pipeline}): 

\begin{enumerate}[label=\textbf{DR\arabic*}, leftmargin=3em]
    \item \textbf{Accommodate diverse viewing modes and their corresponding dynamics while mitigating mode-specific challenges.} We have labeled the items in Table \ref{formative-study-findings} and addressed them accordingly in section \ref{ui}.
    
    \item \textbf{Reduce the volume of danmaku by excluding low-quality, knowledge-irrelevant comments and consolidating duplicate ones.} 
    In addition to minimizing the distraction of video-viewing, reducing the number of danmaku can increase information density. This allows users to concentrate more on individual comments, lowering the chances of comments scrolling away before being fully read. Furthermore, it is easier to locate specific danmaku, as they are less likely to be overshadowed by others.
    % In addition to increasing information density, reducing the number of danmaku can minimize the distraction to video-viewing. This allowed users to concentrate more on individual comments, reducing the frequency of instances where danmaku would scroll away before being fully read. Furthermore, it made it easier to locate specific danmaku, as they were less likely to be overshadowed by others.
    
    \item \textbf{Clearly present the diverse themes of danmaku and the structures of collective knowledge.} The various perspectives danmaku offers can enhance the sense of discussion and provide viewers with more sources of information. Further dissecting video knowledge and arranging danmaku around the corresponding knowledge points can highlight their inherent connections and the overall information structure.
    % Further organizing the patterns of danmaku with the video content would clarify the information structure and strengthen the correspondence between them.
\end{enumerate}

% \yh{While the correspondence of DRs and Findings is obvious to me, but I am not sure if it is the case for readers. Can I just simply list the DRs here?}

% \begin{table}[h!]
% \centering
% \begin{tabular}{| m{6cm} | c | c | c |}
%   \hline
%   \rowcolor{lightgray}
%   \textbf{Challenge} & \textbf{Overview} & \textbf{Focused} & \textbf{Exploration} \\
%   \hline
%   Mutual distractions  & \ding{55} & \ding{51} & \ding{55} \\
%   \hline
%   Short duration & \ding{55} & \ding{51} & \ding{55} \\
%   \hline
%   Challenges in navigating desired danmaku & \ding{51} & \ding{55} & \ding{55} \\
%   \hline
%   \textbf{Low Information Density} & \ding{51} & \ding{51} & \ding{55} \\
%   \hline
%   \textbf{Obscure Information Structure} & \ding{51} & \ding{51} & \ding{55} \\
%   \hline
% \end{tabular}
% \caption{Challenges in different modes}
% \end{table}

\section{Design}
% Driven by the design requirements from the \hyperref[formative-study]{Formative Study}, this section began with a thematic analysis to explore the patterns of collective knowledge in online science videos. Based on the insights and data derived from the analysis, we proposed an NLP pipeline to process danmaku, based on which we implemented CoKnowledge, an interactive system to support viewers' assimilation of collective knowledge in science videos.

To meet the aforementioned design requirements, we must ground our proposed features in an empirical understanding of the danmaku content. We thus started with a content analysis to explore the patterns of collective knowledge in online science videos with time-synced comments. Based on the insights and labeled data, we proposed a natural language processing (NLP) pipeline to automatically filter, classify, cluster danmaku, and map them to the most relevant video segments. On top of the pipeline, we designed CoKnowledge, an interactive system to support viewers' assimilation of collective knowledge in science videos.

\subsection{Content Analysis}
\label{content-analysis}
\subsubsection{Data Collection}
We collected the data for our content analysis from Bilibili's "Knowledge Zone" channel. 
% \xm{"section" may be confused with paper section..} 
% , where 45\% of the videos are reported to pertain to knowledge sharing \cite{bilibili2021}. 
According to \cite{zhang2023understanding, he2024engage}, scientific knowledge videos can be categorized into seven distinct domains based on the knowledge-related hashtags in this channel: \textit{health and medical science, life sciences, earth and space sciences, social sciences and arts, mathematics and physics, computer science}, and \textit{chemistry and material sciences}.
% This section features a total of 33 hashtags designed to organize scientific knowledge. Following \cite{zhang2023understanding, he2024engage}, seven distinct domains encompassing all 33 hashtags were identified for scientific knowledge videos: \textit{health and medical science, life sciences, earth and space sciences, social sciences and arts, mathematics and physics, computer science, and chemistry} and \textit{material sciences}.
To ensure comprehensive topic coverage, we employed the official Bilibili API to crawl five videos and their corresponding danmaku. The selected videos adhered to the following criteria: (1) a minimum duration of three minutes to guarantee sufficient content, (2) over 200 danmaku comments to enhance the likelihood of capturing substantial collective knowledge, and (3) a production date within the last five years (2019 or later) to reflect current trends in science video production. The data collection occurred on February 21, 2024, yielding 28,088 danmaku comments for the 35 videos. Detailed information on each video and its corresponding danmaku is provided in the supplementary material. 
% \xm{how many danmaku in total for the 35 videos did you gather? Say that detailed information about every video and its danmaku is presented in the supplementary material...}

\subsubsection{Content Analysis Pipeline}
% Following the video selection, two authors manually coded 3,000 randomly sampled danmaku comments to explore their informative potential. 
Two authors manually coded 3,000 danmaku comments randomly sampled from the dataset to explore their informative potential.
% For each comment, they assigned a code based on a joint analysis of the video context, which involved locating the timestamp of the danmaku's appearance and watching a brief segment to understand the context of the danmaku.
Initially, the coders independently reviewed 300 samples to gain a preliminary understanding of danmaku quality. Through discussion, they defined \textit{`knowledge danmaku'} as live comments that facilitate the construction or digestion of collective knowledge. This definition encompassed danmaku which inherently presented knowledge as well as promoted the understanding of other knowledge. 
% Subsequently, they independently labeled the first 300 samples to assess inter-rater reliability. 
Subsequently, they independently assigned one-hot \textit{`knowledge danmaku'} labels to the first 300 samples. For each comment, they located the timestamp of its appearance and watched a brief video segment to determine its relevance to the presented knowledge. 
% Utilizing Cohen’s kappa \cite{tavakol2011making}, a statistical measure to quantify agreement among coders, the two coders achieved a high inter-rater reliability score of 0.96. 
The two coders achieved a high inter-rater reliability score of 0.96 measured in Cohen’s kappa \cite{tavakol2011making}. 
Following this validation, they resolved the conflicts and coded the remaining 2,700 samples, ultimately identifying 1,251 \textit{knowledge danmaku} from the entire set.

% \xm{Be sure these are the correct steps...} 
We further adopted an integrated deductive and inductive approach to conducting a content analysis on the \textit{knowledge danmaku} pool regarding their information themes. 
% The deductive aspect draws on the foundational work of Wu \cite{wu2019danmaku}and He \cite{he2021beyond}, which informed the development of our inductive content analysis. 
More specifically, we deductively started with predefined codes from existing literature on Wu \cite{wu2019danmaku} and He \cite{he2021beyond} and then inductively developed new ones that may emerge from the context of science videos. 
% This inductive component allows for the possibility of revising existing themes and incorporating new themes that may emerge from the context of science videos \cite{hadi2022gamification}.
After several rounds of discussion and iterations on the existing codes as sifting through 1000 data samples, the two coders reached a consensus on the final codebook, achieving saturation. They identified five dominant categories of \textit{knowledge danmaku} with respect to the video content (details presented in Table \ref{information-theme}).
After finalizing the codebook, the coders independently labeled 100 samples to validate inter-rater reliability, achieving a Cohen’s kappa of 0.90. With such verification, they proceeded to code the remaining samples with theme labels, which were subsequently used for building classifiers. 
% \xm{Did you also count the occurrence of each category? If so, you may want to add the information to Table 2. This is related to your later claim that interpretation is the biggest category...} 

\begin{table*}[h!]
\centering
\begin{tabular}{p{3.4cm} p{4cm} p{5cm} p{1.2cm}}
\hline
Information category     & Definition                                                                                                         & Example         & Frequency                                                                                                                                                                                                       \\
\hline
Interpretation          & Articulating personal opinions and analyses of the science video content.                   & \textit{Russia invests heavily in scientists; bringing over one Euler would recoup all the expenses.}      & 54.2\%                                                                                                                    \\\hline
Inquiry                 & Formulating questions pertaining to the content of the science video.                       & \textit{Why is the probability of irrational numbers equal to 1?}    & 13.2\%                                                                                                                                                           \\\hline
Experience sharing      & Contributing personal experiences that relate to the themes addressed in the science video. & \textit{My mother passed away from this disease, which was discovered to be liver metastasis. I hope that one day the world can eradicate cancer.}     & 11.8\%                                                                         \\\hline
Concept noting          & Identifying and emphasizing concepts relevant to the content of the science video.          & \textit{D'Alembert's criterion.}          & 11.6\%                                                                                                                                                                                      \\\hline
Supplementary knowledge & Providing additional background information to complement the video knowledge.              & \textit{"Han" refers to a geographical location; its original meaning pertains to the Han River, which later extended to denote the regions through which the Han River flows, and subsequently acquired additional meanings.}  & 9.3\%
\\
\hline
\end{tabular}
\caption{Danmaku's information categories with regard to science video content, along with their definitions, examples, and frequencies in our labeled samples.}
\label{information-theme}
\end{table*}

\subsubsection{Verification of Danmaku's Distinctive Value}
To verify danmaku's potential for facilitating fine-grained knowledge co-construction in online science videos compared to traditional comments, we randomly selected one video from each of seven domains within the video pool and collected their transcript, traditional comments, and danmaku. Utilizing Named Entity Recognition (NER), we identified entities in the video transcripts and labeled whether they were addressed by danmaku or traditional comments. An entity was considered addressed if it met either of the following criteria: (1) an exact match appeared in the danmaku or comments, or (2) the entity was referenced through an abbreviation, alias, synonym, or alternative expression in the danmaku or comments.
Entity coverage rates for danmaku and traditional comments were statistically compared using the Wilcoxon Signed Rank test \cite{woolson2005wilcoxon}. The results revealed that danmaku achieved significantly higher entity coverage than traditional comments (\textit{Z} = -2.34, \textit{p} = 0.016, \textit{Eff. size} = 0.894). Detailed data is provided in the supplementary material. The findings confirm that, compared to traditional comments, danmaku contributes more nuanced and comprehensive augmentations to specific and detailed aspects of video content.

\subsection{Danmaku Processing}

To automate the extraction of \textit{knowledge danmaku} and label their information themes (\textbf{DR2, DR3}), we fine-tuned a \textit{Llama-2 Chinese} model \cite{cui2023efficient} with a \textit{LoRA} adapter \cite{hu2021lora} using the 3,000 annotated danmaku samples. A six-class classifier was developed to determine whether a specific danmaku qualified as \textit{knowledge danmaku} and, if so, to identify its corresponding information themes. We conducted a 7:3 split for training and testing the samples, resulting in a good performance with an F1-score of 72.8\%. This demonstrated that the fine-tuned classifier effectively processed danmaku from all science video topics.
% \xm{[R1 3rd comment: shall we mention that one fine-tuned model can be used for all classes/topics of science videos?]}
 
The next step is to determine the stance of danmaku; according to the \hyperref[formative-study]{Formative Study} participants, clarifying viewpoints in comments can enhance the sense of co-presence and mitigate knowledge bias (\textbf{DR3}). 
% clarifying the stance of danmaku can enhance the sense of co-presence and mitigate knowledge bias (\textbf{DR3}). 
% The content analysis revealed that the interpretations were the most numerous and exhibited the greatest variability in stance, while other themes had relatively consistent attitudes. 
Our content analysis revealed that \textit{interpretation} comments took up the biggest portion of \textit{knowledge danmaku} and exhibited the greatest variability in stance, while those in other categories showed relatively consistent attitudes.
Therefore, we further analyzed the stance of each \textit{interpretation} comment as positive, neutral, or negative by inputting them into a widely recognized sentiment analysis model, \textit{cardifnlp/twitterxlm-roberta-base-sentiment} \cite{barbieri2021xlm, wolf2020transformers} developed by \textit{HuggingFace} \cite{huggingface}. Consequently, we identified a total of seven categories for \textit{knowledge danmaku}. 
% To further streamline the danmaku comments, 
\begin{figure*}[h]
  \centering
  \includegraphics[width=0.8\linewidth]{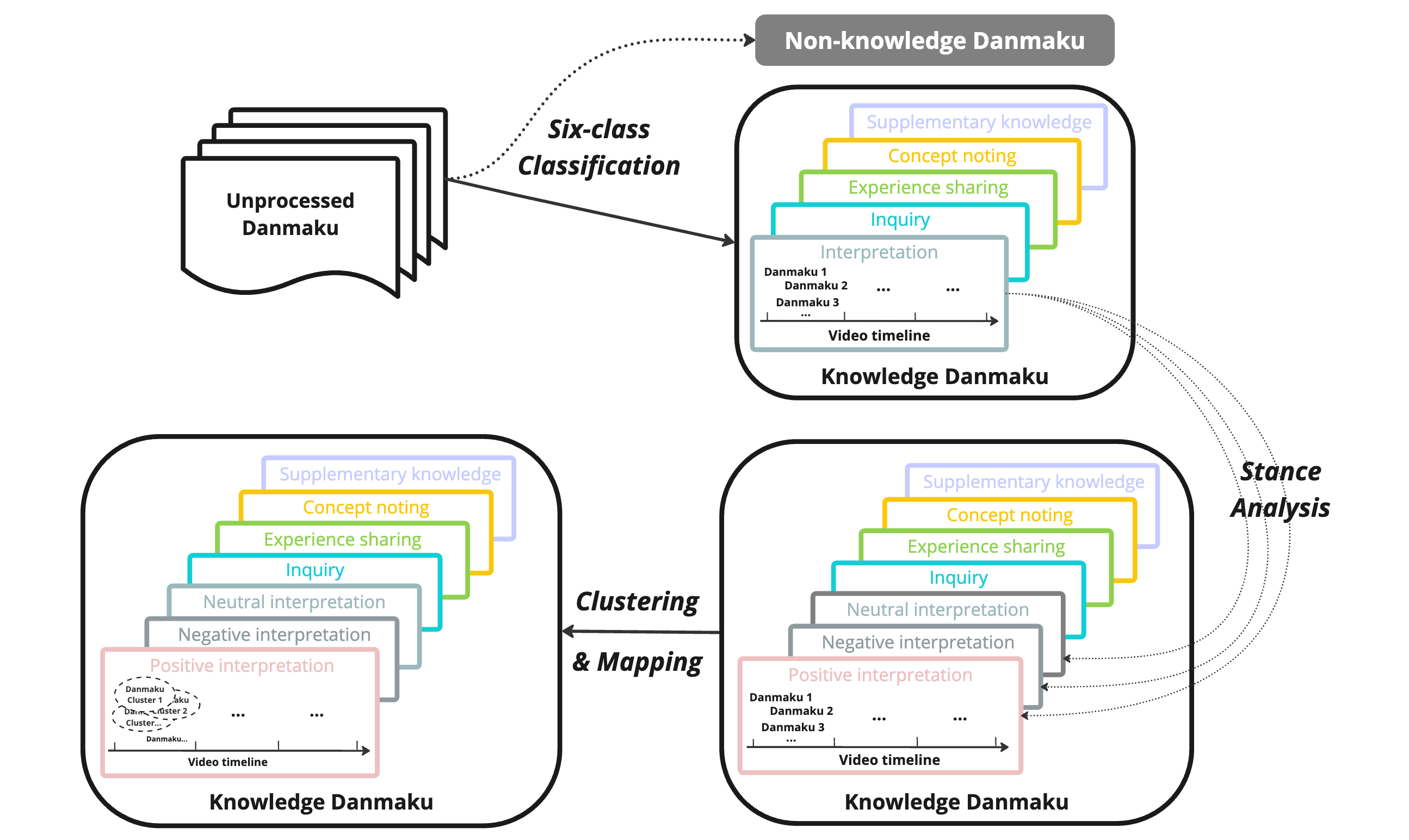}
  \caption{Danmaku processing pipeline for CoKnowledge. }\label{fig:danmaku-processing}
  \Description{Danmaku processing pipeline for CoKnowledge.}
\end{figure*}

To further reduce repetitive information, we utilized \textit{Density-Based Spatial Clustering of Applications with Noise} (DBSCAN) \cite{ester1996density} to group semantically similar danmaku within the same theme and time segment (\textbf{DR3}). The centroid of each cluster was designated as the representative danmaku.
% , which, scaled by the cluster's comment count, replaced all comments within the cluster \xm{what does the second half mean? I think it is more relevant to the user interface part...}. 
Concerning the potential delays of danmaku relative to the video timestamps they are intended to anchor \cite{ma2017video}, we utilized GPT-4 \cite{achiam2023gpt} to map processed danmaku to their corresponding segments based on semantic similarity and temporal proximity. The specific algorithms and prompts are detailed in the supplementary material.
These steps collectively formed our automated NLP pipeline for danmaku processing (Figure \ref{fig:danmaku-processing}), which supports the operation of CoKnowledge.

\subsection{User Interface}
\label{ui}

To meet the design requirements specified in section \hyperref[drs]{3.3} and Figure \ref{fig:design-pipeline}, we developed CoKnowledge to facilitate users' assimilation of collective knowledge. 
% Reflecting the viewing modes identified in the \hyperref[formative-study]{Fomrative Study}, 
Following people's ways of viewing science videos with danamku as identified in the \hyperref[formative-study]{Formative Study}, 
CoKnowledge comprised three distinct modes of displaying collective knowledge, namely, overview mode, focused mode, and exploration mode. Each mode was equipped with specific design elements tailored to address the unique needs and challenges encountered by users within that mode (\textbf{DR1}). Additionally, we previously observed that different modes were associated with particular video-watching behaviors (Table \ref{formative-study-findings}); hence, CoKnowledge enabled automatic mode transitions based on user behaviors to create a seamless viewing experience.
The following subsections detail 
% \xm{use present tense when describing what this paper is about / section writes about...} 
the key user interface (UI) elements of each mode, specifying which items (Table \ref{formative-study-findings}) 
% \xm{??? this is very confusing -- unclear what you mean by features. In Fig 1, the features are the modes and NLP functions...} 
they were designed to address.
We utilized widely recognized tools for the tasks and conducted performance evaluations. All technical details (e.g., algorithms, prompts, evaluations) are included in the supplementary material.

\subsubsection{Overview Mode}
\label{overview-mode}

\begin{figure*}[h]
  \centering
  \includegraphics[width=0.9\linewidth]{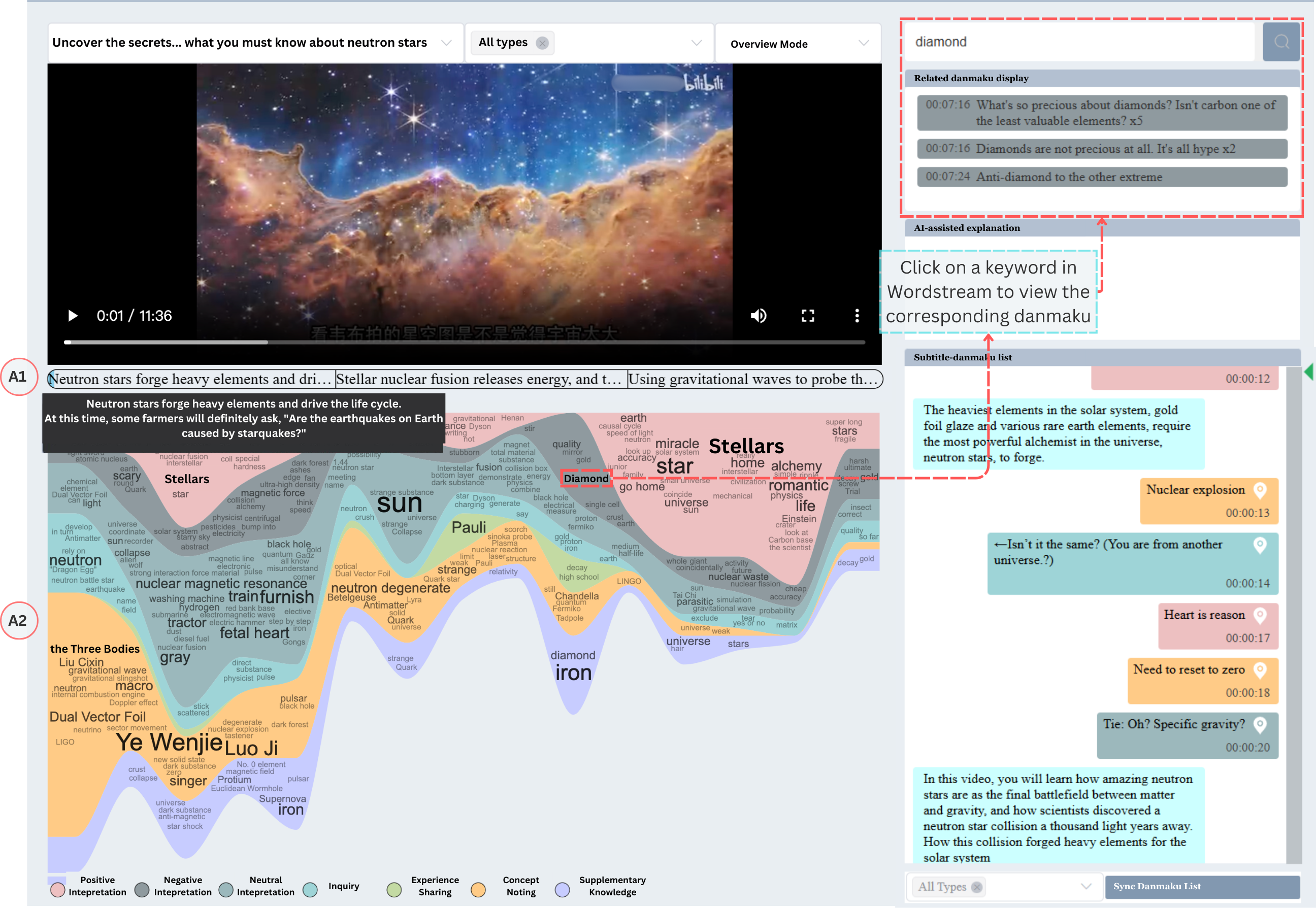}
  \caption{Interface of Overview Mode. A1: Progress bar directory. A2: Wordstream with legend filter. The explanation of each feature is in section \hyperref[overview-mode]{4.3.1}.}\label{fig:overview}
  \Description{Interface of Overview Mode. The explanation of each feature is in section 4.3.1.}
\end{figure*}

% When users initially loaded the video and had not yet begun playback, 
The system defaulted to
% \xm{I think you need to add `the' before the three modes as a specific reference}
\textbf{Overview Mode} when users just loaded a video (Figure \ref{fig:overview}).
This mode included a progress bar directory (Figure \ref{fig:overview} A-1) and a Wordstream (Figure \ref{fig:overview} A-2). 
These components functioned in tandem to synchronously display video and danmaku knowledge, respectively. This setup enabled users to identify specific timestamps in the video while quickly obtaining an outline of the collective knowledge presented.

In particular, the progress bar directory offered a segmented summarization of the video knowledge (Table \ref{formative-study-findings} \circled{1}\circled{4}\circled{10}) by dividing the whole video into several semantically coherent sections and condensing the content in each into a sentence using Bilibili's official AI video summary tool \cite{bilibili_video}.
% \xm{state the tool/algorithm you use. If it is LLM, say that the prompts are included in the supplementary materials..} 
% This directory corresponded with the video’s progress bar, 
This directory aligned with the video timeline,
with the length of each section reflecting the duration of the corresponding video segment. When users hovered over a specific position in the progress bar directory, the system displayed the video transcript at that timestamp, enabling users to access more concrete content if they found the summaries too abstract.

WordStream, positioned below the progress bar directory, combined a stacked area chart with word clouds. Inspired by \cite{heimerl2015citerivers, nguyen2022wordstream, liu2012tiara}, such representation not only maintained an aesthetic appeal \cite{heimerl2015citerivers} but also effectively handled and analyzed large amounts of qualitative, time-series, and topic-based data \cite{nguyen2022wordstream, liu2012tiara}, making it suitable for rendering an overview of \textit{knowledge danmaku} throughout the video. 
Other design alternatives, such as heatmaps and line charts, cannot capture the rich semantic information conveyed by danmaku.
The x-axis of Wordstream encoded time, while the y-axis encoded the quantity of danmaku at specific timestamps, with colors of the bands indicating the categories of the \textit{knowledge danmaku}. 
The word clouds contained keywords extracted from the danmaku appearing at each timestamp.
We used GPT-4 \cite{achiam2023gpt} to identify a representative word for each comment.  
% The word clouds were generated from keywords extracted from the danmaku appearing at each timestamp, with GPT-4 \cite{achiam2023gpt} being utilized to identify a representative keyword for each comment. 
The larger a keyword was displayed, the more danmaku containing that keyword appeared at that timestamp.

\begin{figure*}[h]
    \centering
    \begin{subfigure}{0.482\textwidth}
        \centering
        \includegraphics[width=\textwidth]{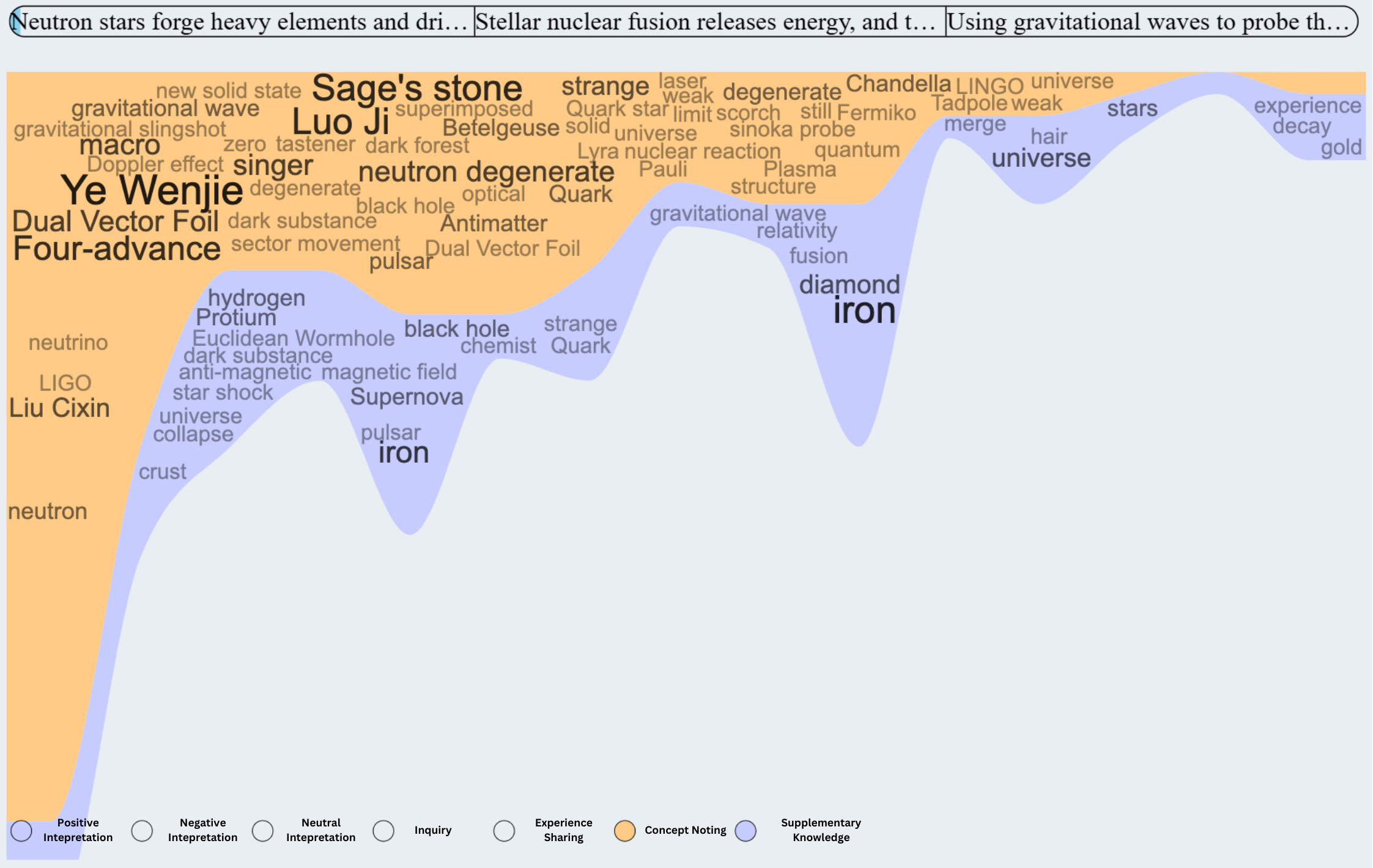}
        \caption{}
        % \caption{Wordstream after filtering for the categories of \textit{concept noting} and \textit{supplementary knowledge}.}
        \label{fig:subfigure1}
    \end{subfigure}
    \hfill
    \begin{subfigure}{0.505\textwidth}
        \centering
        \includegraphics[width=\textwidth]{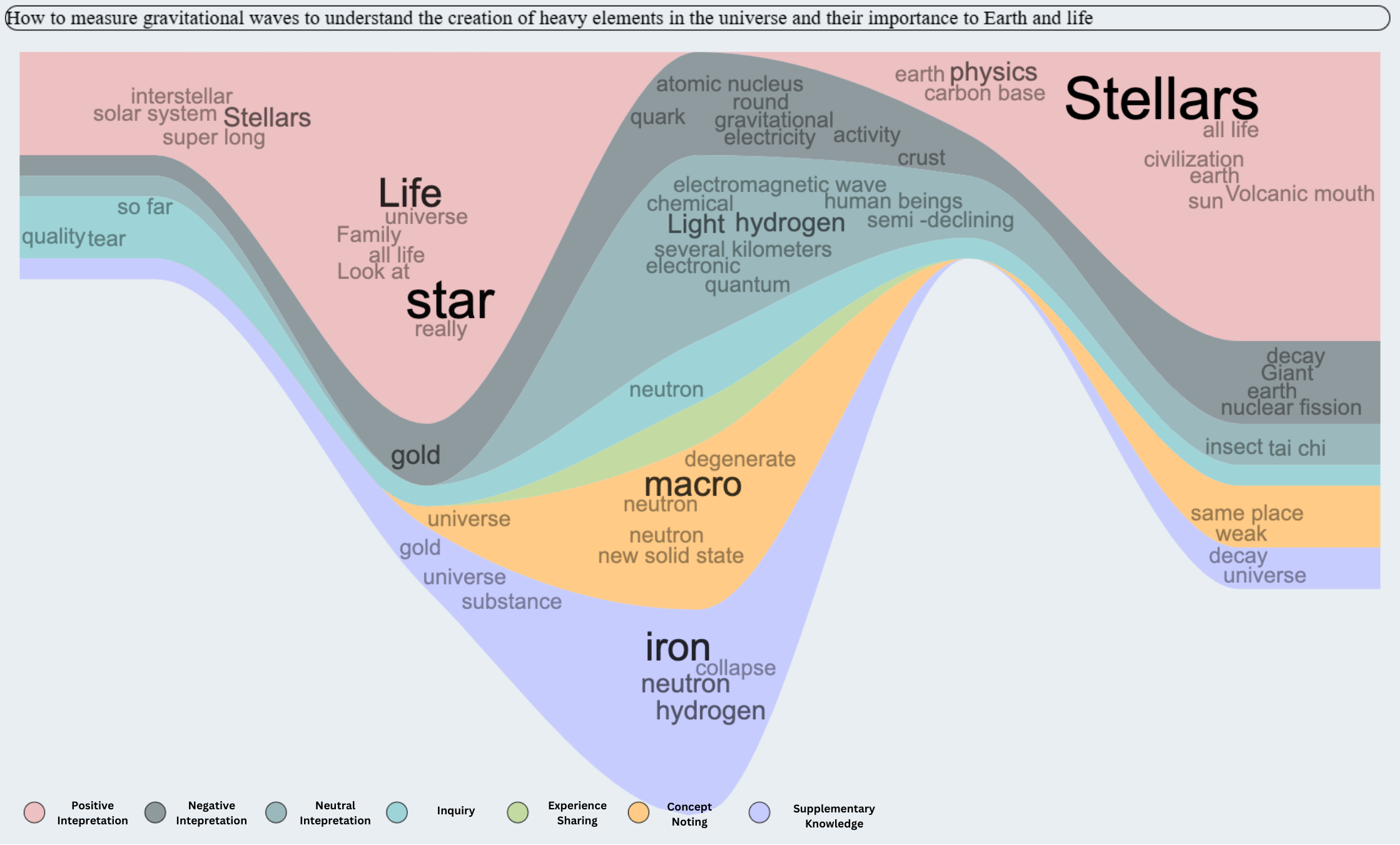}
        \caption{}
        % \caption{An expanded segment of Wordstream and progress bar directory.}
        \label{fig:subfigure2}
    \end{subfigure}
    \caption{More interactions of \hyperref[overview-mode]{Overview Mode}: a) Wordstream after filtering for the categories of \textit{concept noting} and \textit{supplementary knowledge}. b) An expanded segment of Wordstream and progress bar directory.}
    \label{fig:combined}
    \Description{More interactions of Overview Mode: a) Wordstream after filtering for the categories of concept noting and supplementary knowledge. b) An expanded segment of Wordstream and progress bar directory.}
\end{figure*}

% This dual representation vividly illustrated the semantic and theme proportions in danmaku at each timestamp while also showcasing their evolution over the course of the video (Table \ref{formative-study-findings} \circled{1}\circled{4}\circled{10}\circled{13}\circled{18}). 
The Wordstream provided a vivid overview of the distribution and gist of different types of \textit{knowledge danmaku} and their evolution over the course of the video (Table \ref{formative-study-findings} \circled{1}\circled{4}\circled{10}\circled{13}\circled{18}). 
Following Shneiderman's visualization mantra \cite{shneiderman2003eyes}, we also enable ``zoom and filter then details-on-demand'' via interactions (Table \ref{formative-study-findings} \circled{7}). 
Users could filter the Wordstream by specifying the categories of interest using the legend below (Figure \ref{fig:subfigure1}). If they were particularly interested in the collective knowledge within a certain video section, they could click on the corresponding segment in the progress bar directory to enlarge it (Figure \ref{fig:subfigure2}). This would cause the segment to expand and display a more detailed summary, and the Wordstream would zoom in accordingly. 
% , with Wordstream updating to reflect only that section.
Moreover, users could click on a keyword in the Wordstream to view the corresponding comments in the side view for detailed exploration (Figure \ref{fig:overview}).
% \xm{say that this part is explained later in subsection xxx}. 

% Users can click on a keyword in the Wordstream to view the corresponding comments in the side view for more detailed exploration (Table \ref{formative-study-findings} \circled{7}). Furthermore, users could filter the Wordstream by specific themes of danmaku using the legend below (Figure \ref{fig:subfigure1}). If users were particularly interested in the collective knowledge of a specific time segment, they could click on the corresponding segment in the progress bar directory to enlarge it (Figure \ref{fig:subfigure2}). This would cause the segment to expand and display a more detailed summary, with Wordstream updating to reflect only that section.

% The progress bar directory and Wordstream functioned in tandem to synchronously display both video knowledge and danmaku knowledge. This setup enabled users to identify specific timestamps in the video while quickly obtaining an outline of the entire collective knowledge presented.

\subsubsection{Focused Mode}
\label{focused-mode}

\begin{figure*}[h]
  \centering
  \includegraphics[width=0.9\linewidth]{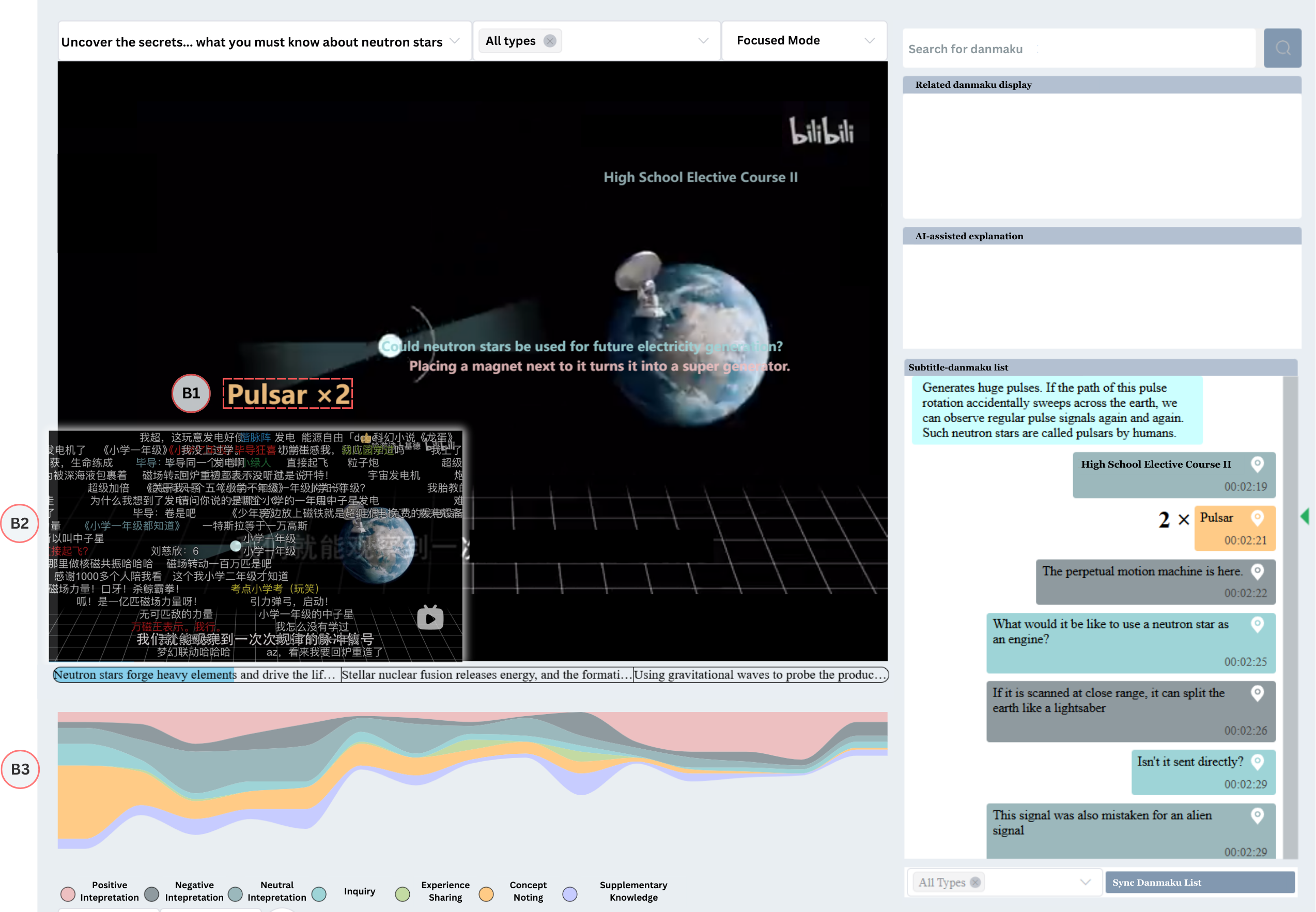}
  \caption{Interface of Focused Mode. B1: Representative danmaku comment with a notation. B2: Unprocessed danmaku stream. B3: Simplified version of the features in Overview Mode. The explanation of each feature is in section \hyperref[focused-mode]{4.3.2}\label{fig:focused}}
  \Description{Interface of Focused Mode. The explanation of each feature is in section 4.3.2}
\end{figure*}

When users began playing the video, CoKnowledge automatically transitioned to \textbf{Focused Mode} (Figure \ref{fig:focused})
% \xm{[Address the position of the figures... they are too far away from where they are first referenced...e.g., Fig 6 is 2 pages behind...]}
, which enlarged the video viewing window (Table \ref{formative-study-findings} \circled{14}) with minimum interactive features (Table \ref{formative-study-findings} \circled{8}). While other modes offered alternative representations of danmaku for digesting collective knowledge, \textbf{Focused Mode} retained the synchronized scrolling of consolidated \textit{knowledge danmaku} across the video to maintain its inherent alignment and contextual relevance to the content. By mitigating danmaku's issues of visual clutter and short display duration in the video view and significantly reducing information load in other views (Table \ref{formative-study-findings} \circled{16}\circled{17}), this mode enabled users to concentrate on both the video and the floating danmaku,
% as they normally do on an original danmaku platform. This design aimed to
delivering an authentic, immersive and streamlined viewing experience (Table \ref{formative-study-findings} \circled{5}\circled{11}).
%This design enabled users to concentrate on both the video and the danmaku, ensuring an immersive and streamlined viewing experience (Table \ref{formative-study-findings} \circled{5}\circled{11}).}

% This design aimed to enhance users’ concentration on the video and floating danmaku (Table \ref{formative-study-findings} \circled{5}\circled{11}). 
To be more specific, as illustrated in Figure \ref{fig:focused}, the number of floating comments significantly decreased with the help of our NLP pipeline (Table \ref{formative-study-findings} \circled{2}\circled{16}\circled{17}). Users could further select the type(s) of danmaku they wish to view on the video screen. To address the issue of short display duration (Table \ref{formative-study-findings} \circled{17}), we implemented the following adjustments. First, longer danmaku floated more slowly. Second, if a comment represented a cluster of similar danmaku, a notation indicating the number of danmaku in the cluster was added (Figure \ref{fig:focused} B-1).
% \xm{why is the notation related to duration?}
As participants in the \hyperref[formative-study]{Formative Study} generally recognized the value of danmaku supported by many users, the larger the number, the bigger the comment appeared, and the more slowly it scrolled.
% the larger the font size and the slower the floating speed was.
Below the video window was a simplified version of the Wordstream (Figure \ref{fig:focused} B-3) from \hyperref[overview-mode]{Overview Mode}, with the word clouds removed and the stacked area chart compressed vertically. It was intended to offer danmaku evolution patterns while minimizing distractions.
% The two features at the bottom served as simplified versions (Figure \ref{fig:teaser} B-1) of \hyperref[overview-mode]{Overview Mode} to minimize distractions.

\subsubsection{Exploration Mode}
\label{exploration-mode}
\begin{figure*}[h]
  \centering
  \includegraphics[width=0.9\linewidth]{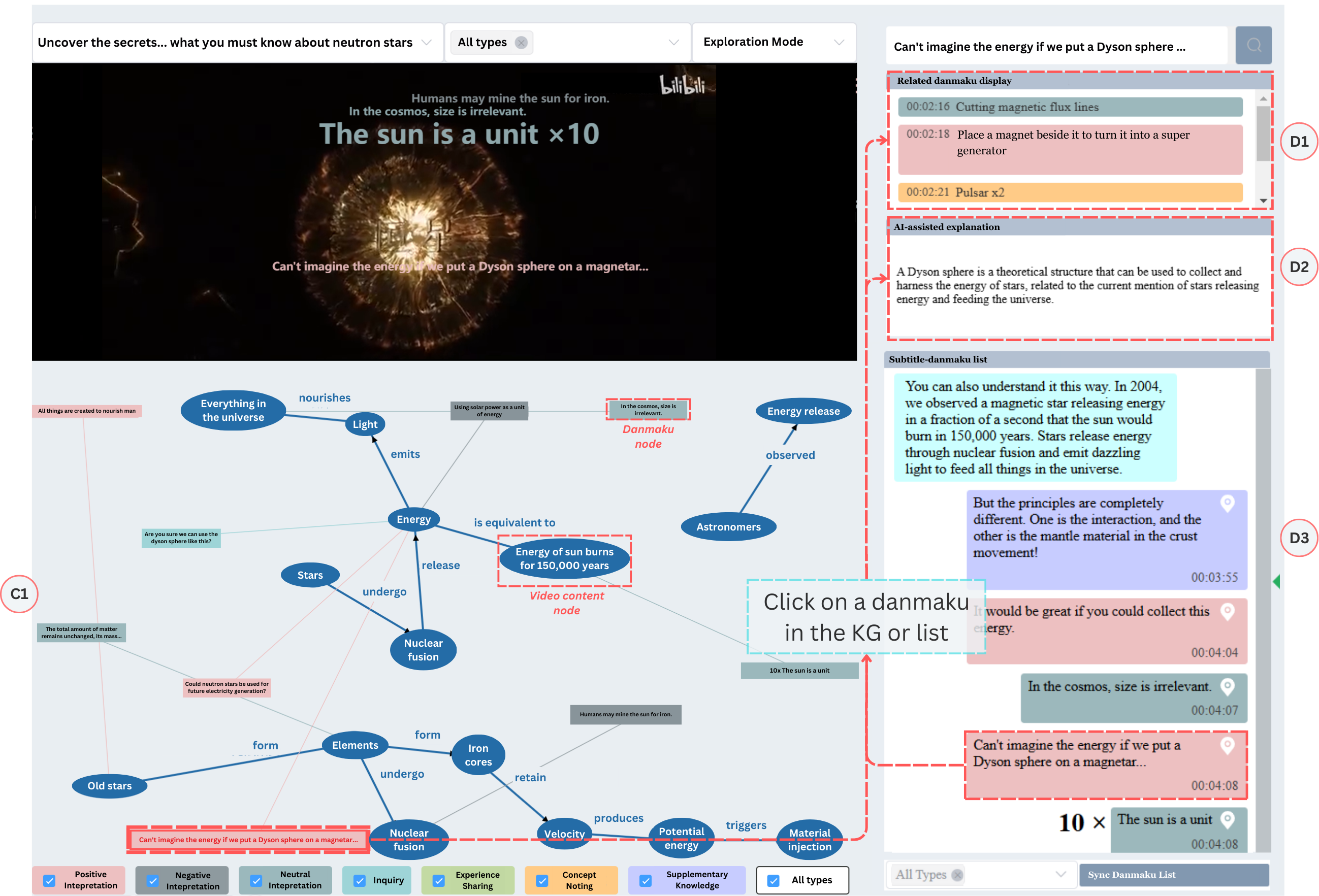}
  \caption{Interface of Exploration Mode. C1: Knowledge graph with legend filter. D1: Related danmaku display. D2: AI-assisted explanation. D3: Subtitle-danmaku list. The explanation of each feature is in sections \hyperref[exploration-mode] {4.3.3} and \hyperref[side-view]{4.3.4}.}\label{fig:exploration}
  \Description{Interface of Exploration Mode. The explanation of each feature is in sections 4.3.3 and 4.3.4.}
\end{figure*}

When users paused at a specific timestamp, CoKnowledge entered \textbf{Exploration Mode} (Figure \ref{fig:exploration}) automatically, displaying the structured collective knowledge within that 20-second segment in the form of a Knowledge Graph (KG) 
% \xm{[How is the knowledge graph constructed? We need to expand the following sentences to highlight that part of the KG comes from video captions/subtitles -- The question by R1]}
below the video window (Table \ref{formative-study-findings} \circled{3}\circled{6}). The blue elliptical nodes and their links within the KG represented video content entities and their relationships. They were extracted from the video transcripts during that period through named entity recognition and relation extraction tasks using GPT-4 \cite{achiam2023gpt} 
% \xm{providing prompt in the supplementary material}
. The square node(s) linked to an elliptical video node represented its most relevant danmaku comment(s), with colors denoting the danmaku category. 
The 20-second segment length was determined through multiple rounds of testing, considering factors such as the number of danmaku, space utilization, graph clarity, and users’ cognitive load.
% Surrounding these nodes were colored square nodes representing the different themes of danmaku comments, with each danmaku attached to the most relevant video node. 
Similar to the other modes, users could filter the categories of danmaku using the legend. When they clicked on a specific square danmaku node, the side view displayed its related danmaku within the short time interval and AI-assisted explanations to facilitate a thorough analysis (Table \ref{formative-study-findings} \circled{9}\circled{12}).

% \xm{I think you should have a separate technical evaluation subsection to show the performances of all the automatic components -- e.g., NLP pipeline, progress bar directory summarization, knowledge extraction, explanation generation, etc.}
% % The 20-second segment length was determined through multiple rounds of testing, considering factors such as the number of danmaku, space utilization, graph clarity, and users’ cognitive load.
% The extraction of triples \xm{unclear what triples refer to...} for the KG was performed using GPT-4 \cite{achiam2023gpt} \xm{providing prompt in the supplementary material}. KG accuracy was defined as the proportion of triples within the KG being correct \cite{gao2019efficient}. A triple was deemed correct if the entities appeared in the video content and their relationships aligned with the narrative. To evaluate GPT-4's performance, we tuned the prompts and randomly selected 35 twenty-second segments from the content analysis video pool for GPT-4 to extract triples. Using manual annotation as the ground truth, GPT-4 achieved a satisfactory accuracy of 85.7\%.

% Similarly, users could filter the themes of danmaku using the legend. Users could also click on specific danmaku to highlight it, with the side view displaying related danmaku and AI-assisted explanations to facilitate a thorough analysis (Table \ref{formative-study-findings} \circled{9}\circled{12}). 

% \subsubsection{Other features}
\subsubsection{Side View}
\label{side-view}

In addition to the automatic mode transitions based on user behavior, users could manually switch modes. The side view (Figure \ref{fig:teaser} D) remained a constant feature that did not change with mode transitions. 

The bottom section of the side view was the subtitle-danmaku list (Figure \ref{fig:exploration} D-3), which displayed video subtitles (aligned to the left) and time-synced comments (aligned to the right) in chronological order. This list automatically synchronized and scrolled with the video playback. 
% Users could click a button below the list to manually align it with the video timestamps \xm{unclear... isn't it synced already?} (Figure \ref{fig:teaser} D-4). 
Users could filter danmaku categories within the list.
The upper sections of the side view were reserved for ``detail-on-demand,'' presenting related danmaku (Figure \ref{fig:exploration} D-1) and AI-assisted explanation (Figure \ref{fig:exploration} D-2) when users clicked on a danmaku node in KG or a comment in the list below. 
% related danmaku display (Figure \ref{fig:teaser} D-1) and AI-assisted explanation (Figure \ref{fig:teaser} D-2), which presented supplementary features when users clicked on a danmaku comment from the KG or the list below. 
% The related danmaku section displayed comments logically associated with the selected danmaku, while the AI-assisted explanation offered insights from GPT-4 \cite{achiam2023gpt} regarding the relationship between the selected danmaku and the video content. 
The related comments, posted within 15 seconds before or after the selected danmaku and logically linked to it, were identified through named entity recognition and entity linking tasks performed by GPT-4 \cite{achiam2023gpt}.
% \yh{To determine this, we first performed named entity recognition (NER) to extract entities from all comments within 20 seconds before and after the selected danmaku, followed by entity linking. A comment was identified as related if it contained an entity linked to any entity of the selected danmaku}.
% \xm{how are they determined?}
The explanations were generated by GPT-4 \cite{achiam2023gpt} to offer insights into the relationship between the selected danmaku and the video content. 
% As mentioned in the \hyperref[overview-mode]{Overview Mode}, the related danmaku display has an extra feature: clicking a Wordstream keyword shows the danmaku comments that formed it at that time.

\section{Evaluation}

This section outlines the detailed experimental configuration, including the setup of conditions, study procedures, and evaluation metrics. We implemented a mixed-methods, within-subject study in which participants viewed science videos with CoKnowledge and a baseline system.

\subsection{Conditions and Data Selection}
% Data/Content Selection/Criteria
We developed a baseline system that retained only three functions from Bilibili \cite{bilibili_about_us}: video playback, danmaku playback, and a danmaku list, excluding all other information, such as traditional comments. This system was designed to closely resemble CoKnowledge in UI components and style to minimize potential confounding variables, while removing any data processing of danmaku implemented in CoKnowledge.

\begin{table*}[h]
\centering
\begin{tabular}{c l l c c l}
\hline
\multicolumn{1}{c}{\textbf{Domain}} & \textbf{Topic}               & \textbf{Length}  & \textbf{Word Count} & \textbf{\textit{Knowledge danmaku}} & \textbf{View}  \\ \hline
\multirow{2}{*}{Health}    & Lactose Intolerance & 13'31'' & 3679                           & 361                                   & 5382k \\ \cline{2-6} 
                           & Alkaline Foods      & 13'49'' & 3598                           & 339                                   & 3103k \\ \hline
\multirow{2}{*}{Astronomy} & Neutron Star        & 11'37'' & 3765                           & 403                                   & 1949k \\ \cline{2-6} 
                           & Saturn              & 12'05'' & 3827                           & 379                                   & 1286k \\ \hline
\end{tabular}
\caption{Information about the selected videos, including the topics, length, transcript word count, number of \textit{knowledge danmaku}, and video views.}
\label{video}
\end{table*}

To demonstrate the generalizability of CoKnowledge, we selected two pairs of videos (as detailed in Table \ref{video}) that meet the following criteria: 1) They span different domains—astronomy and health. 2) They feature different narrative orientations—one introductory and the other debunking. 3) They vary in content complexity—one is more technically rigorous, while the other is more relatable to everyday life. In the subsequent content, we refer to these pairs as astronomy and health videos. Each video exceeds ten minutes, includes over 2,000 danmaku comments, and has over one million views, indicating both popularity and a substantial information load,  making them typical cases for CoKnowledge to address.
To ensure comparability, the videos within each pair were selected from the same domain, had analogous information loads (measured by video length, transcript word count, and the number of \textit{knowledge danmaku}), and were sourced from the same content creator to maintain consistency in communication style. Additionally, two authors meticulously reviewed the videos to confirm no overlap in the knowledge covered within each pair.

\subsection{Participants and Procedure}
To ensure the effectiveness of statistical analysis with sufficient scope, 
we performed an \textit{a priori} power analysis using G*Power \cite{faul2007g} to determine the required sample size. The analysis indicated that a minimum of 20 participants was needed to detect large effects (d = 0.8) with 90\% power in a Wilcoxon signed-rank test, assuming an alpha level of 0.05. After obtaining institutional IRB approval, we ultimately recruited 24 participants (10 female, 14 male, aged 18-31, with backgrounds in business, engineering, arts, and science) through online advertisements, social media, and word-of-mouth. 
Participants were eligible for the study if they were frequent viewers of science videos on platforms featuring danmaku and self-reported low prior knowledge of the experiment's video topics. 
% Demographic details are summarized in Table \ref{demographics}. 
We excluded participants from our previous study to avoid potential bias.

\begin{figure*}[h]
  \centering
  \includegraphics[width=0.8\linewidth]{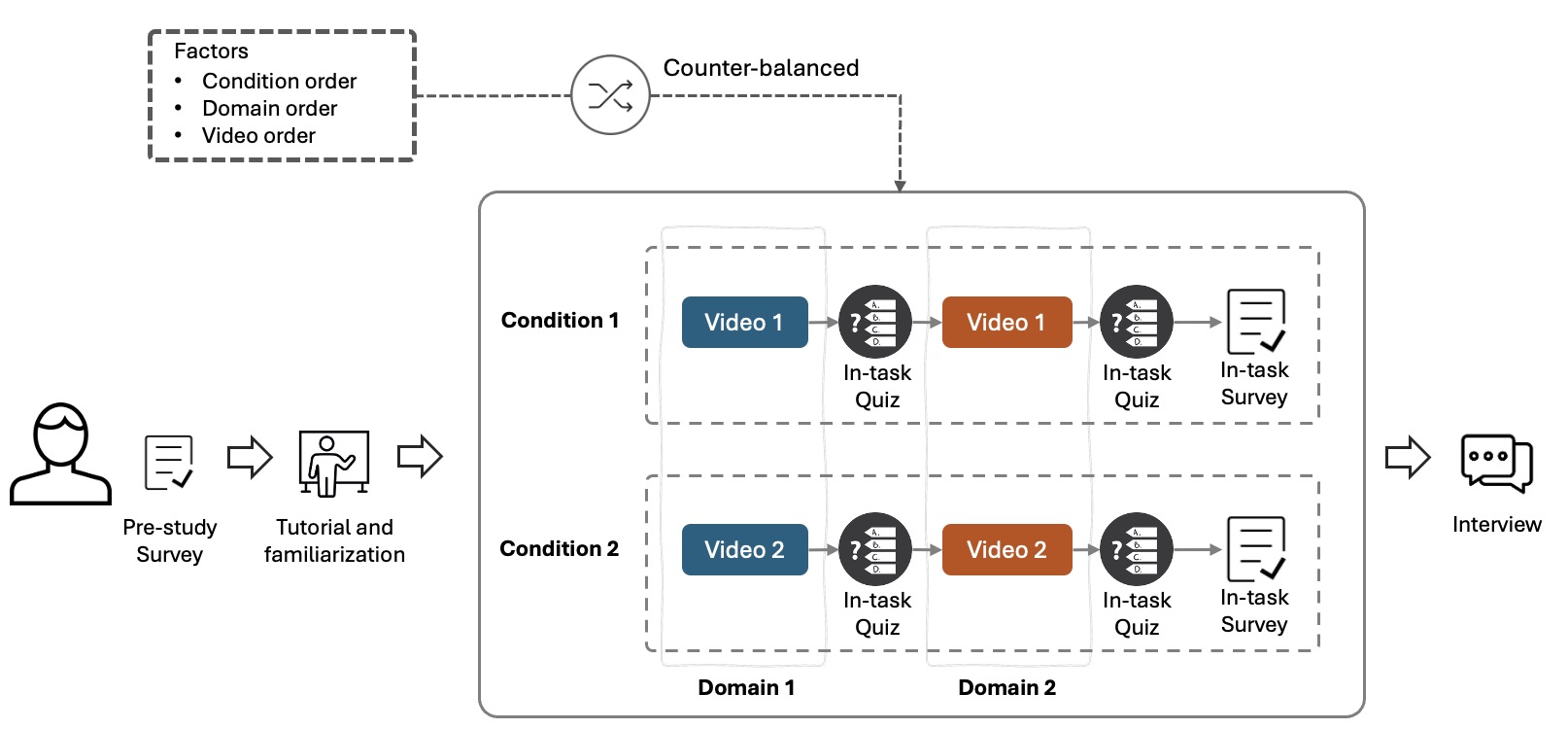}
  \caption{User study procedure: Participants took part in a within-subject study, watching two video pairs (each from a different domain) using two different systems in a counter-balanced order. The study included a pre-study survey, in-task quizzes, and an interview.}\label{fig:study-procedure}
  \Description{User study procedure: Participants took part in a within-subject study, watching two video pairs (each from a different domain) using two different systems in a counter-balanced order. The study included a pre-study survey, in-task quizzes, and an interview.}
\end{figure*}

Figure \ref{fig:study-procedure} illustrates the experimental procedure. Prior to the study, participants completed a pre-study survey to gather demographic information and confirm their eligibility. During the experiment, with participants' consent, we recorded their screen activity, audio, and collected system logs.  Each participant began with a 10-minute tutorial, during which they could freely explore CoKnowledge and raise questions. Following the tutorial, participants were briefed on the context and tasks. There were a total of four tasks: two using CoKnowledge, and two using the baseline system. For each task, participants were given 16 minutes to watch a video and 10 minutes to complete a quiz. The order of task combinations was counterbalanced.
% \begin{itemize}[noitemsep, topsep=0pt]
%     \item Neutron Star; Alkaline foods (Baseline) - Saturn; Lactose intolerance (CoKnowledge)
%     \item Saturn; Lactose intolerance (Baseline) - Neutron Star; Alkaline foods (CoKnowledge)
%     \item Saturn; Lactose intolerance (CoKnowledge) - Neutron Star; Alkaline foods (Baseline)
%     \item Neutron Star; Alkaline foods (CoKnowledge) - Saturn; Lactose intolerance (Baseline)
% \end{itemize}
After each condition, participants were asked to fill out an in-task survey to provide feedback on the system. The study concluded with a semi-structured interview that explored participants' experiences with CoKnowledge, their use of specific features and underlying behavioral rationale,  and a comparison with the baseline system.
% , as well as systems' utilization of danmaku. 
The study lasted approximately 120 minutes, and each participant received compensation aligned with the local hourly wage standard.

\subsection{Measurements}
Following the established evaluation pipeline for interactive systems \cite{weibelzahl2020evaluation, xia2022persua}, we assessed the systems across three key dimensions: system usefulness, design \& interaction, and system usability. All in-task surveys used a 7-point Likert scale, with specific questions detailed in the supplementary material.
% We reviewed relevant literature to construct the in-task surveys, with all questions utilizing a 7-point Likert scale. \yh{The specific questions of in-task surveys are presented in the supplementary material.}

\subsubsection*{System Usefulness}

According to Bloom’s taxonomy \cite{bloom1971taxonomy}, knowledge (the ability to recall learned information) and comprehension (the ability to interpret meanings and concepts) are the fundamental levels of learning outcomes. Based on this framework, we designed the in-task quizzes using a single-choice question (SCQ) format, structured into four sections to assess participants' understanding: comprehension of video knowledge, recall of video knowledge, comprehension of danmaku knowledge, and recall of danmaku knowledge. 
% \yh{Based on Bloom’s taxonomy \cite{bloom1971taxonomy}, which identifies knowledge (recall of information) and comprehension (interpretation of concepts) as fundamental learning outcomes, we designed in-task quizzes using a single-choice question (SCQ) format to assess participants' understanding. Quizzes were divided into four sections: comprehension and recall of both video and danmaku knowledge.}
To maintain consistent quiz difficulty within each domain, we first created an extensive question pool for each video and conducted a pilot test with 9 participants (with inclusion criteria identical to those in the formal user study). These participants watched each video under baseline conditions and answered the questions in the pool. We then scored their responses and selected subsets of questions from the pool as the formal study quiz, ensuring that the average score for quiz questions in each section was equivalent across the two videos within the same domain. The final quiz included 16 SCQs—6 for video knowledge and 10 for danmaku knowledge—with an option to select 'Not sure.' To increase the reliability of the learning outcome measurements, we discouraged participants from guessing by penalizing incorrect responses: -1/3 points for wrong answers, +1 point for correct answers, and 0 points for selecting the 'Not sure' option.

To evaluate CoKnowledge's effectiveness in supporting users' assimilation of collective knowledge, we conducted a multi-level comparison of in-task quiz scores across the two conditions. We also gauged participants' confidence by analyzing the number of quiz questions where "Not sure" was not selected. In addition to these objective measures, participants also provided self-assessments of perceived comprehension, recall, confidence, and efficiency in the in-task surveys.
Furthermore, echoing the \hyperref[formative-study]{Formative Study}, we aimed to evaluate CoKnowledge's ability to harness the benefits of danmaku while mitigating its potential drawbacks. To achieve this, we gathered users' perceptions on the relevant aspects in the in-task surveys.%(as detailed in Table \ref{subjective})%.

\subsubsection*{Design \& Interaction}

% Our goal was to understand how users interacted with CoKnowledge and perceived the helpfulness of its features. We asked participants to rate the helpfulness of different modes, data processing methods, and UI components of CoKnowledge in the in-task survey. We also reviewed the recorded videos and system logs to analyze participants' behavior patterns with CoKnowledge.
We investigated CoKnowledge's feature helpfulness by asking participants to rate different modes, data processing methods, and UI components in the in-task surveys. We also analyzed recorded videos and system logs to identify behavior patterns during their interactions with CoKnowledge.

\subsubsection*{System Usability}

Balancing functionality and usability is a common challenge for computer-supported systems \cite{kang2021metamap,goodwin1987functionality}. In our evaluation, we first measured task workload using the NASA Task Load Index \cite{hart1988development}, followed by an assessment of overall usability using a shortened version of the System Usability Scale (SUS) questionnaire \cite{brooke1996sus}.

% \begin{table}[h]
% \centering
% \begin{tabular}{cc}
% \hline
% \textbf{Category} & \textbf{Questions} \\
% \hline
% \multirow{4}{*}{Perceived performance} 
% & \textit{I found myself comprehending the collective knowledge completely.} \\
% & \textit{I found myself able to recall the collective knowledge accurately.} \\
% & \textit{I felt confident in assimilating the collective knowledge.} \\
% & \textit{I am effective in assimilating the collective knowledge.} \\
% \hline
% \multirow{3}{*}{Danmaku advantage}     
% & \textit{I felt as though we were watching videos with others.} \\
% & \textit{Danmaku helped me avoid the knowledge bias with the video uploader.} \\
% & \textit{I felt danmaku is closely related to the video content.} \\
% \hline
% \multirow{5}{*}{Danmaku drawback}      
% & \textit{I felt that danmaku and video content interfered with each other.} \\
% & \textit{I often missed danmaku because it disappeared too quickly.} \\
% & \textit{I could easily locate a specific danmaku comment.} \\
% & \textit{I found the information density of danmaku to be low.} \\
% & \textit{I perceived the information structure of the collective knowledge to be clear.} \\
% \hline
% \end{tabular}
% \caption{Survey questions to subjectively evaluate system usefulness.}
% \label{subjective}
% \end{table}
\section{Results}

% All participant ratings mentioned in Section 5.4 were measured using a 7-point Likert scale. For the data items that were measured under both conditions, we first conducted the Shapiro-Wilk test \cite{shapiro1965analysis} to assess their normality. The results indicated that most of the items did not meet the normality assumption. Therefore, we chose to utilize the Wilcoxon Signed Rank test \cite{woolson2005wilcoxon}. 
In this section, we present our evaluation of CoKnowledge, focusing on its usefulness, design \& interaction, and usability. 
According to Shapiro-Wilk test \cite{shapiro1965analysis}, most of the measurement items did not meet the normality assumption. 
Therefore, following \cite{sakai2017probability, sullivan2012using, tomczak2014need}, we applied the Wilcoxon Signed Rank test \cite{woolson2005wilcoxon} and reported the test statistic, p-value, and the corresponding effect sizes. Descriptive statistics were used to summarize participants' responses regarding the perceived helpfulness of CoKnowledge's features.
Our findings are supported by statistical inferences and further enriched by qualitative reflections from the participants.

\subsection{System Usefulness}
\label{sec:usefulness}
We evaluated CoKnowledge's effectiveness in supporting the assimilation of collective knowledge while sufficiently leveraging the capabilities of danmaku. Detailed statistics are presented in Table \ref{Usefulness}. 
% \xm{The table is at the end of the paper. Move it here.}.

\begin{table*}[]
\begin{tabular}{ccccccc}
\hline
Category                                                                           & Item                                & \begin{tabular}[c]{@{}c@{}}Baseline\\ Mean (S.D.)\end{tabular} & \begin{tabular}[c]{@{}c@{}}CoKnowledge\\ Mean (S.D.)\end{tabular} & Z              & p        & Eff. Size   \\ \hline
\multicolumn{1}{c}{}                                                               & Total                               & 4.95(2.76)                             & 7.73(2.67)                               & -4.66    & 0.000*** & 0.82   \\
\multicolumn{1}{c}{}                                                               & Danmaku                             & 1.79(1.79)                             & 3.90(1.99)                               & -4.81   & 0.000*** & 0.86   \\
\multicolumn{1}{c}{}                                                               & Video                               & 3.16(1.91)                             & 3.83(1.54)                               & -2.07  & 0.039*  & 0.30   \\
\multicolumn{1}{c}{}                                                               & Danmaku comprehension               & 1.08(1.29)                             & 2.35(1.43)                               & -3.81    & 0.000*** & 0.55   \\
\multicolumn{1}{c}{}                                                               & Danmaku recall                      & 0.72(1.30)                             & 1.56(1.23)                               & -2.57     & 0.010**  & 0.37   \\
\multicolumn{1}{c}{}                                                               & Video comprehension                 & 1.48(1.28)                             & 1.88(1.09)                               & -1.99      & 0.047*  & 0.29   \\
\multicolumn{1}{c}{\multirow{-7}{*}{Grade}}                                        & Video recall                        & 1.68(1.04)                             & 1.94(0.88)                               & -1.39      & $0.164^{-}$  & 0.20   \\ \hline
                                                                                   & Total                               & 12.06(3.35)                            & 12.92(3.33)                              & -2.01       & 0.044*  & 0.29   \\
                                                                                   & Danmaku                             & 6.54(2.91)                             & 7.38(2.78)                               & -2.24     & 0.025*  & 0.32   \\
                                                                                   & Video                               & 5.52(0.88)                             & 5.54(0.82)                               & -0.10       & $0.918^{-}$  & 0.01   \\
                                                                                   & Danmaku comprehension               & 3.25(1.63)                             & 3.81(1.45)                               & -2.47      & 0.014*  & 0.36   \\
                                                                                   & Danmaku recall                      & 3.29(1.53)                             & 3.56(1.44)                               & -1.20     & $0.228^{-}$  & 0.17   \\
                                                                                   & Video comprehension                 & 2.77(0.47)                             & 2.83(0.38)                               & -0.73     & $0.467^{-}$  & 0.11   \\
\multirow{-7}{*}{\begin{tabular}[c]{@{}c@{}}Confidence\\ (Objective)\end{tabular}} & Video recall                        & 2.75(0.56)                             & 2.71(0.58)                               & 0.41     & $0.685^{-}$  & 0.06   \\ \hline
                                                                                   & Comprehension                       & 3.96(1.63)                             & 4.74(1.60)                               & -1.86     & 0.049*  &  0.38      \\
                                                                                   & Recall                              & 3.83(1.40)                             & 4.63(1.38)                               & -1.98     & 0.047*  & 0.40       \\
                                                                                   & Confidence                          & 3.71(1.55)                             & 4.83(1.20)                               & -2.51   & 0.012** & 0.62   \\
\multirow{-4}{*}{\begin{tabular}[c]{@{}c@{}}Subjective\\ Evaluation\end{tabular}}  & Efficiency                          & 4.13(1.60)                             & 5.13(1.33)                               & -2.26    & 0.024*  & 0.46   \\ \hline
                                                                                   & Co-presence                         & 3.38(1.74)                             & 4.25(1.77)                               & -2.31      & 0.021*  & 0.47   \\
                                                                                   & Avoid knowledge bias                 & 3.92(1.67)                             & 5.00(1.45)                               & -2.69     & 0.007** & 0.55   \\
\multirow{-3}{*}{\begin{tabular}[c]{@{}c@{}}Danmaku\\ Advantage\end{tabular}}      & Close connection with video content & 4.04(1.85)                             & 5.58(1.38)                               & -3.57    & 0.000***& 0.73   \\ \hline
                                                                                   & Mutual distractions $\downarrow$              & 5.13(1.03)                             & 4.42(1.67)                               & 1.96      & $0.051^{+}$  & 0.40   \\
                                                                                   & Short display duration    $\downarrow$                  & 4.71(1.60)                             & 3.92(1.72)                               &  1.67  & $0.096 ^{+}$  & 0.38   \\
                                                                                   & Danmaku locating difficulties    & 3.92(1.64)                             & 4.67(1.44)                               & -1.45     & $0.147^{-}$  & 0.30   \\
                                                                                   & Low information density    $\downarrow$         & 5.83(1.01)                             & 3.63(1.69)                               & 3.82    & 0.000***& 0.78   \\
\multirow{-5}{*}{Challenges}                                                 & Obscure information structure       & 3.83(1.69)                             & 5.29(1.27)                               & -2.89   & 0.004** & 0.59   \\ \hline
\end{tabular}
\caption{The statistical analysis of system usefulness with Baseline and CoKnowledge, where the p-value (-: p > .100, +: .050 < p < .100, *:p < .050, **:p < .010, ***:p < .001) is reported. $\downarrow$  indicates that a lower score is better. By default, higher scores are better.}
\label{Usefulness}
\end{table*}

\subsubsection{Collective Knowledge Assimilation} 
While using CoKnowledge, we observed a significant improvement in the overall scores of in-task quizzes compared to the baseline system with an average of 56.1\% increase. %The average score across the entire test increased by 56.1\% with CoKnowledge. 
A more detailed analysis revealed 
% \xm{present or past tense? Be consistent across the paper.} 
that the quiz scores for both the danmaku segment and the video segment were significantly higher under the system condition than the baseline condition. Specifically, the mean score increased by 117.8\% for the danmaku segment and 21.1\% for the video segment with CoKnowledge. 
Further examination showed a significant difference in danmaku comprehension, danmaku recall, and video comprehension, although not in video recall. The participants also had significantly fewer `unsure' answers in danmaku-related questions. This indicates that participants acquired more knowledge with high confidence from the danmaku. Meanwhile, their mastery of video content was not undermined. 

These findings align with the subjective evaluations, where participants reported significantly higher levels of \textbf{comprehension and recall} of collective knowledge, higher \textbf{efficiency} in knowledge digestion, and stronger \textbf{confidence} in answering the quiz when using CoKnowledge (Table \ref{Usefulness}).
% In the post-study interview, the participants reflected on the perceived advantages of CoKnowledge. 
% \xm{Give the list of participants (PXX, PXX, ...) holding each view.}
In the post-study interview, all participants recognized CoKnowledge's remarkable effectiveness in aiding collective knowledge assimilation.
P14 stated that \textit{``after using CoKnowledge, I feel like my knowledge has significantly expanded and developed.''} 
% \xm{use `` '' instead of ```` for quotes...}
P19 explained the considerable improvement in absorbing knowledge from live comments, \textit{``Initially, I could only capture less than 10\% of the danmaku [normally], but with CoKnowledge, I was able to process most of the danmaku. As for the video, I could remember most of it even with the baseline system, so the improvement with CoKnowledge wasn’t as significant.''}
Still, several participants noted the benefits of understanding the videos brought by our system. For instance, \textit{“The reduction in the quantity of danmaku [after data processing] allowed me to focus on the relationship between the danmaku and the video... [When answering questions in the video segment,] I first recalled some danmaku and then the corresponding video content''} (P15).

In addition to the overall video analysis, we also analyzed quiz scores separately for different videos, with the statistics presented in Table \ref{Video_type}.
% \xm{move the table here}. 
%Briefly, for the hard science videos, all items exhibited significant differences across the two conditions except for the grade related to video comprehension. For the popular science videos, we observed significant differences in the overall test score ($Z = 2.96, p = 0.003$), danmaku segment ($Z = 3.81, p < 0.001$), and danmaku comprehension segment ($Z = 2.84, p = 0.005$), while there were no significant differences in the video segment score, danmaku recall segment, video comprehension segment, or video recall segment.
Overall, the system's effect on enhancing the assimilation of astronomy video content is greater than that of health videos. 
% , which can be attributed to the lower difficulty in understanding the content of popular science videos and their associated danmaku. 
Several participants (P1-3, P12, P13) gave the reason for this phenomenon in the interview, for example, \textit{``the health videos are more relatable to daily life and their danmaku is rarely difficult to understand. Even without the system, my knowledge absorption was quite effective''} (P1).
% As P1 noted: \textit{ ``[In terms of content,] the popular science videos are more relatable to daily life, especially the danmaku, which are rarely difficult to understand. Even without the system, my knowledge absorption was quite effective.``}

\begin{table*}[]
\begin{tabular}{ccccccc}
\hline
Category                    & Item                    & \begin{tabular}[c]{@{}c@{}}Baseline\\ Mean (S.D.)\end{tabular} & \begin{tabular}[c]{@{}c@{}}CoKnowledge\\ Mean (S.D.)\end{tabular} & \multicolumn{1}{c}{Z}  & \multicolumn{1}{c}{p}   & Eff. Size    \\
\hline
                            & Total                   & 3.64(2.70)                                                     & 6.94(2.13)                                                       & -3.65 & 0.000*** & 0.75 \\
                            & Danmaku                 & 1.54(1.93)                                                     & 3.68(1.96)                                                       & -3.05 & 0.002** & 0.62 \\
                            & Video                   & 2.10(1.88)                                                     & 3.26(1.32)                                                       & -2.01 & 0.044* & 0.41 \\
                            & Danmaku comprehension & 0.93(1.45)                                                     & 2.08(1.19)                                                       & -2.74 & 0.006** & 0.56 \\
                            & Danmaku recall          & 0.61(1.15)                                                     & 1.60(1.27)                                                       & -2.07 & 0.038* & 0.42 \\
                            & Video comprehension     & 0.81(1.14)                                                     & 1.35(0.96)                                                       & -1.59 & $0.111^{-}$ & 0.33 \\
\multirow{-7}{*}{Astronomy} & Video recall            & 1.29(1.11)                                                     & 1.92(0.84)                                                       & 2.37 & 0.018* & 0.48 \\
\hline
                            & Total                   & 6.26(2.15)                                                     & 8.51(2.97)                                                       & -2.96 & 0.003** & 0.60 \\
                            & Danmaku                 & 2.04(1.65)                                                     & 4.13(2.04)                                                       & -3.81 & 0.000*** & 0.78 \\
                            & Video                   & 4.22(1.25)                                                     & 4.39(1.56)                                                       & -0.89  & $0.372^{-}$ & 0.18 \\
                            & Danmaku comprehension & 1.22(1.11)                                                     & 2.61(1.61)                                                       & -2.84 & 0.005** & 0.58 \\
                            & Danmaku recall          & 0.82(1.46)                                                     & 1.51(1.22)                                                       & -1.51  & $0.131^{-}$ & 0.31 \\
                            & Video comprehension     & 2.15(1.04)                                                     & 2.42(0.94)                                                       & -1.13 & $0.260^{-}$ & 0.23 \\
\multirow{-7}{*}{Health}   & Video recall            & 2.07(0.82)                                                     & 1.97(0.94)                                                       & 0.16 & $0.876^{-}$ & 0.03 \\
\hline
\end{tabular}
\caption{The statistical analysis of quiz grades for different video pairs with Baseline and CoKnowledge, where the p-value (-: p > .100, +: .050 < p < .100, *:p < .050, **:p < .010, ***:p < .001) is reported. For all items, a higher score indicates better performance.}
\label{Video_type}
\end{table*}

\subsubsection{Danmaku Utilization}
\label{sec:danmaku-util}

This subsection examines whether CoKnowledge can fully harness danmaku's potential  (Table \ref{Usefulness}), as specified in the requirements derived from the \hyperref[formative-study]{Formative Study}. 
% \xm{reference the corresponding subsection}. %as indicated by findings from the formative study. 

Participants reported a significantly higher \textbf{information density} in the danmaku when using CoKnowledge, and all participants highly commended this aspect in the interview. For instance, P16 noted that \textit{``This greatly lowered the barrier to grasping collective knowledge.''} Furthermore, while using CoKnowledge, participants found the \textbf{structure of collective knowledge} to be significantly clearer and the \textbf{connection between the danmaku and the video content} to be significantly tighter. They attributed this clarity to the integration of various features in the interview. To be more specific, P2, P5, P7, and P15 noted that the classification of danmaku made \textit{``the structure of the danmaku (with respect to the video content) less chaotic''} (P2). Additionally, the progress bar directory provided an overall organization of the video content (P15), and the knowledge graph (P9, P13, P14) enhanced \textit{``the clarity of the structure of danmaku and the corresponding video segments''} (P13).

Even though danmaku comments appear in multiple locations in our system, the participants did not find it significantly more \textbf{challenging to locate} them than using the baseline.
We also observed a marginally significant reduction in the reported \textbf{mutual distraction between danmaku and video content} and in the likelihood of missing danmaku due to \textbf{short display durations} under the CoKnowledge condition. P1, P6, P14, and P16 highly praised the setting that allowed longer danmaku to remain on screen for an extended period, despite that some found the coloring of floating comments (for encoding classification results) somewhat distracting (e.g., P10). 
% Consequently, the participants perceived a significantly stronger \textbf{connection between the danmaku and the video content}. Several of them (P5, P15) attributed this improvement to the classification of danmaku.

Furthermore, participants reported a significantly enhanced \textbf{sense of co-presence} with CoKnowledge. Some participants (P4-6, P12, P24) mentioned that the classification of danmaku allowed them to \textit{``selectively focus on those that shared common topics (with them), which strengthened the sense of identification''} (P24).  Additionally, \textit{``the clearer perception of differing stances in danmaku content''} increased their \textit{``sense of interaction''} (P4). As a result, the system condition demonstrated a significantly stronger ability to \textbf{mitigate knowledge bias} compared to the baseline. As P5 remarked, \textit{``I deliberately focused on danmaku with negative interpretations rather than blindly trusting the uploader’s narrative.''} 
However, some participants held different views. P20 remarked, \textit{``The system's processing of information places me in a higher position to observe a large group of people’s conversations from a distance.''} %which creates a greater sense of distance.``} 
P23 further added, \textit{``After the data is processed [and abstracted], it loses the original sense of watching together with everyone.''}

% There was only a marginal difference in the reported mutual distraction between danmaku and video content across the two conditions. P10 noted that \textit{``the colorful danmaku was somewhat distracting.``} Similarly, participants perceived only a marginal reduction in missed danmaku with short display durations. However, some participants (P1, P6, P14, P16) highly praised the setting that allowed longer danmaku to remain on screen for an extended period. No significant difference was observed regarding CoKnowledge's effectiveness in facilitating navigation to the desired danmaku.

% Concerning the two primary obstacles, participants reported a significantly higher information density in the danmaku when using CoKnowledge. During the interviews, all participants highly commended CoKnowledge's handling of this aspect. For instance, P16 noted that \textit{``This significantly lowered the barrier to assimilating collective knowledge.``} Furthermore, while using CoKnowledge, participants found the structure of collective knowledge to be significantly clearer. In the interviews, participants attributed this clarity to the integration of various features. They noted that the classification of danmaku made \textit{``the structure of the danmaku (with respect to the video content) less chaotic``} (P2, P7). Additionally, the progress bar directory provided an overall organization of the video content (P15), and the knowledge graph enhanced \textit{``the clarity of the structure of danmaku and the corresponding video segments``} (P9, P13, P14).

\subsection{Design \& Interaction}
\label{sec:interaction}
We examined participants' interactions with CoKnowledge and their perceptions of its feature helpfulness. Figure \ref{fig:use_pattern} presents an overview of usage patterns across different stages. Descriptive statistics of the ratings are presented in Figure \ref{fig:user-rating}.

\subsubsection{Use Patterns}
\label{sec:use-parttern}

\begin{figure*}[h]
  \centering
  \includegraphics[width= \linewidth]{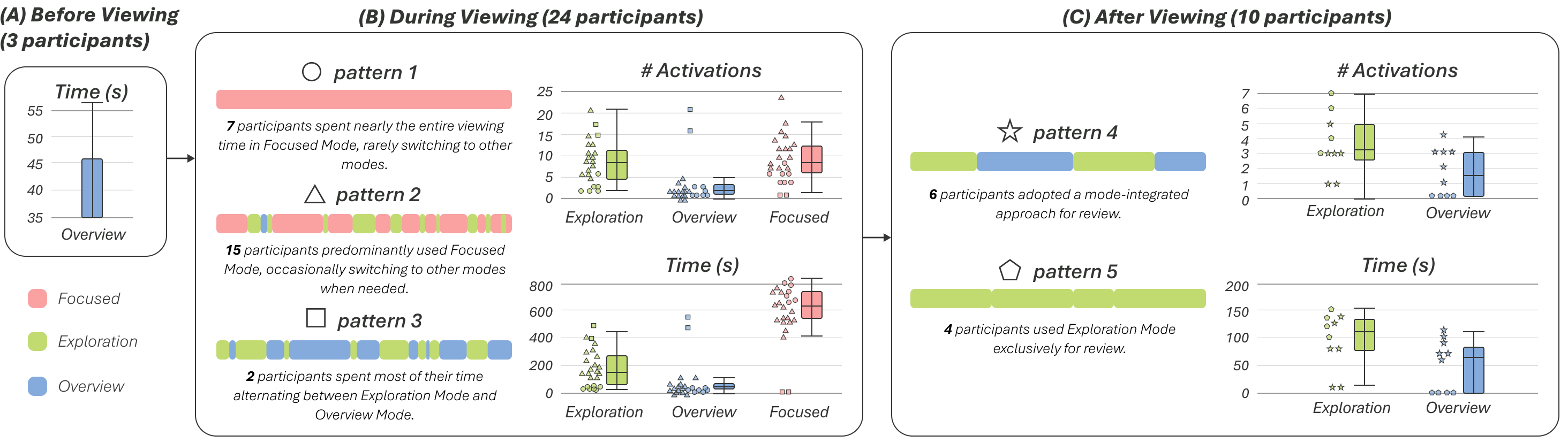}
  \caption{Usage patterns of CoKnowledge across the stages of before, during, and after viewing. Mode transition timelines represent the average mode usage of participants within each pattern group. Box plots illustrate the quantitative distribution of mode activations and time spent in each mode, with data point shapes indicating the pattern associated with individual participants.}\label{fig:use_pattern}
  \Description{Usage patterns of CoKnowledge across the stages of before, during, and after viewing. Mode transition timelines represent the average mode usage of participants within each pattern group. Box plots illustrate the quantitative distribution of mode activations and time spent in each mode, with data point shapes indicating the pattern associated with individual participants. The specific patterns are further explained in section 6.2.1}
\end{figure*}

We identified the participants' interaction patterns with the system by reviewing and analyzing system log data and screen recordings, and further compared them with participants' behaviors in a traditional viewing environment.

Before starting the video (Figure \ref{fig:use_pattern} A), only a few participants (P18, P20, P23) utilized features within the overview mode for preview. Two of them (P18, P23) aimed to capture peaks in danmaku activity and points of interest. P20, on the other hand, sought to \textit{``get a general sense of the video and danmaku to have an overall understanding [of story flow].''} Many participants (16 out of 24) appreciated the functionality of overview mode, emphasizing that, unlike traditional viewing methods relying on brief video descriptions or titles for a cursory understanding of the video (P4, P9), CoKnowledge enabled them to preview the evolution of the entire video and extensive danmaku in a more structured manner.

During video viewing (Figure \ref{fig:use_pattern} B), most participants (22 out of 24) preferred to spend the majority of their time in the focused mode, aligning with traditional viewing habits. For instance, P5 noted, \textit{``My previous habit [of watching science videos] led me to focus primarily on the video section.''} A few of them (P4, P5, P8, P15, P20) preferred to simultaneously view the video and the subtitle-danmaku list on the side view to follow the content in real-time. In contrast to the majority, two participants (P16, P22) spent most of their time in the exploration and overview modes, with P22 stating, \textit{``The amount of information acquired in focused mode is relatively limited.''} Although many participants retained certain aspects of traditional viewing habits, we observed considerable changes in their actual behavior. In conventional environments, participants often minimized danmaku distractions by disabling it (P2, P17, P20), shrinking its display area (P1, P4, P6, P7, P11), or increasing transparency (P1, P4, P19), focusing almost entirely on the video content (P3, P5, P13). With CoKnowledge, the manageable danmaku volume eliminated the need for such adjustments, enabling participants to engage more deeply with danmaku knowledge in their innate format. When encountering confusing or intriguing information during viewing, participants also displayed new strategies. Traditionally, they might have paused, rewatched, or searched online. With CoKnowledge, most participants (15 out of 24) switched to the exploration mode, leveraging features like KG, subtitle-danmaku list, AI-assisted explanations, and related-danmaku display to efficiently investigate the content from alternative perspectives with reduced rewatching. Nonetheless, some participants (P2, P11, P12) found mode switching cumbersome due to the automatic adjustment of video view size, which hindered visibility of video details in exploration mode and overview mode (P11, P22).

After completing the video (Figure \ref{fig:use_pattern} C), some participants (10 out of 24) chose to spend time reviewing the video content. Among them, several (P2, P15, P16, P20, P22, P24) opted to use the overview mode for a comprehensive review and then navigated to the desired timestamps before switching to the exploration mode for thorough analysis (P2, P24). In addition to this mode-integrated approach, other participants (P3, P8, P14) relied on memory to locate specific video timestamps before utilizing the exploration mode, while P6 revisited specific content by browsing the subtitle-danmaku list. Without CoKnowledge, nearly all participants (22 out of 24) indicated they would not have revisited the video, or if they did, it would have been limited to monotonous rewatching.

Overall, compared to traditional viewing environment, CoKnowledge offered diverse features to support various stages of interaction, catering to different needs and accommodating a range of user strategies, thereby enabling multi-dimensional exploration of collective knowledge.

\subsubsection{Helpfulness of the features}
\begin{figure*}[h]
  \centering
  \includegraphics[width=0.85\linewidth]{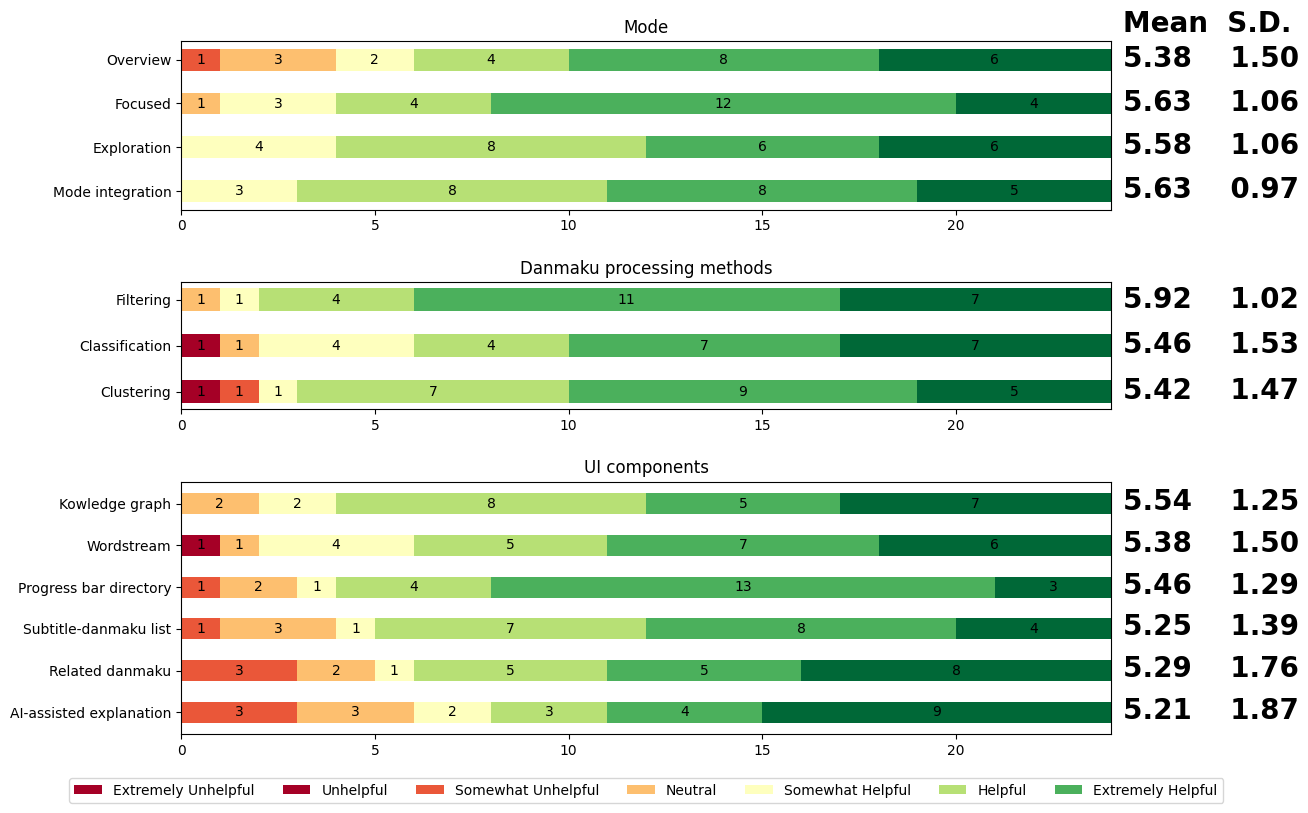}
  \caption{User ratings on CoKnowledge's features, where \textit{UI} stands for \textit{User Interface}. For all items, a higher rating indicates better performance.}
  \label{fig:user-rating}
  \Description{User ratings on CoKnowledge’s features, where UI stands for User Interface. For all items, a higher rating indicates better performance. All features received high ratings with means over 5.}
\end{figure*}

The participants' ratings indicated that the three modes were overall successful in fulfilling the corresponding knowledge needs and were effectively integrated to enhance the assimilation of collective knowledge (Figure \ref{fig:user-rating}). 
We detail the user perception of each design feature below. 
%We begin by discussing the data processing methods. 

\textbf{Filtering} emerged as the CoKnowledge's highest-rated feature. Our participants found the volume of danmaku in the baseline condition overwhelming, leading to an inability to retain any of the comments (P3, P24); by contrast, CoKnowledge's filtering of danmaku reduced the quantity to a manageable level (P9, P11, P14). On the one hand, this allowed participants to absorb the knowledge from time-synced comments while watching the video (P1, P7, P8, P15). 
As P15 stressed, \textit{``Danmaku filtering not only enables me to focus more specifically on \textit{knowledge danmaku} but also more efficiently concentrate on the corresponding video content.''}
On the other hand, this feature significantly lowered their cognitive load when interacting with other features (e.g., \textit{Wordstream} and \textit{KG}) (P15 - 17, P20). However, some (P4, P5, P13) noted that if their goal was pure entertainment, they might avoid filtering, as \textit{``the removed danmaku [, though not informative,] brought joy''} (P13).

The key benefits of \textbf{danmaku classification} were discussed in Section \ref{sec:danmaku-util}. This feature enhanced the sense of co-presence (P12, P24), bridged the gap between the danmaku and video content (P5, P15), and helped structure collective knowledge (P2, P7). Additionally, explicitly labeling the theme of each specific comment allowed users to focus on those related to negative interpretations and supplementary knowledge, thereby avoiding knowledge bias (P5, P15). However, using color to encode these classifications was also noted as potentially distracting (P10).

As for \textbf{danmaku clustering}, many participants (10 out of 24) praised it for reducing the need to mentally process repetitive danmaku while maintaining an idea of the volume. 
% \xm{correct?}.
P2 mentioned, \textit{``When I see a message coming in a large quantity, I pay more attention to it.''} P3 used the information for a different reason, arguing that the same comment posted by many users tended to lack valuable information, and clustering could help avoid such comments all at once. %Regardless of the underlying reasons, this method reduced participants' cognitive load. 
In brief, these data processing methods collectively constitute an NLP pipeline, serving as the foundational framework upon which CoKnowledge is built.

%As shown in Table \ref{User-ratings}, participants rated most of the UI components above 5. 
All main UI components received a user rating above 5 (out of 7), as shown in Figure \ref{fig:user-rating}. 
Regarding \textbf{Wordstream}, some participants noted that it helped them \textit{``understand the content at each time segment, and the classification also provides insights into the overall content of danmaku''} (P1, P19). P4 further commented that the peaks in Wordstream allow them to \textit{``identify moments of intense discussion.''} However, some participants (P2, P3, P6) found Wordstream difficult to understand, as certain keywords were challenging to interpret without contextual information.
Hence, participants' preferences for using Wordstream varied. P23 remarked, \textit{``By reviewing Wordstream before watching the video, I can get a general sense of the danmaku feedback, form some hypotheses, and watch the video with expectations, which can then be validated.''} Some other users (P1, P13, P21, P22) were inclined to employ this feature after watching the video to review the flow of collective knowledge, while P3 referred to it during the video to recall previous content.
The \textbf{progress bar directory}, the other component in overview mode, was favored for its precise indexing of specific parts of the video, besides its ability to structure the video content as mentioned in Section \ref{sec:danmaku-util}. 
% Regarding the other component in overview mode, the progress bar directory, beyond its ability to structure the video content as mentioned in Section \ref{sec:danmaku-util}, P5 noted that it could be used to precisely index specific parts of the video.
 
As for \textbf{KG}, some participants felt that it enabled them to quickly and intuitively grasp the content (P9, P13, P14, P22) and provided a clear structure for the collective knowledge of the current video segment (P15). P17 added, \textit{``With KG, I can specifically understand which aspect of the video content each danmaku corresponds to.''} P12 also mentioned a sense of achievement after fully understanding a knowledge graph. However, others found that the knowledge graph contained too much information for a single segment, making it difficult to keep up (P1, P3, P10). Some also complained that \textit{``the links were cluttered [with many danmaku]''} (P2) and found KG hard to understand due to abstract keywords (P3, P4).

%(May consider delete this paragraph to shorten the result section?) Regarding the features in the side view, 
Approximately one-fourth of the participants considered the \textbf{subtitle-danmaku list} to be a substantial source of information. P9 found it \textit{``very convenient and clear... If I miss something from the danmaku or video content, a quick glance allows me to understand it, making it an excellent tool for filling in the gaps.''}
Some participants (P5, P22) queried \textbf{related danmaku} to \textit{``gain a more comprehensive understanding of a particular comment''} (P22).
Several users (P1, P14, P16, P23) sought out \textbf{AI-assisted explanations}, as they recognized its assistance in quickly grasping unfamiliar concepts (P1, P23), especially for \textit{``hard science videos with challenging terminology''} (P6).

% \xm{Things mentioned in this subsection are mostly pros. Anything to improve or missing?}

\subsection{System Usability}
\label{sec:usability}
\begin{figure*}[h]
  \centering
  \includegraphics[width= 0.85\linewidth]{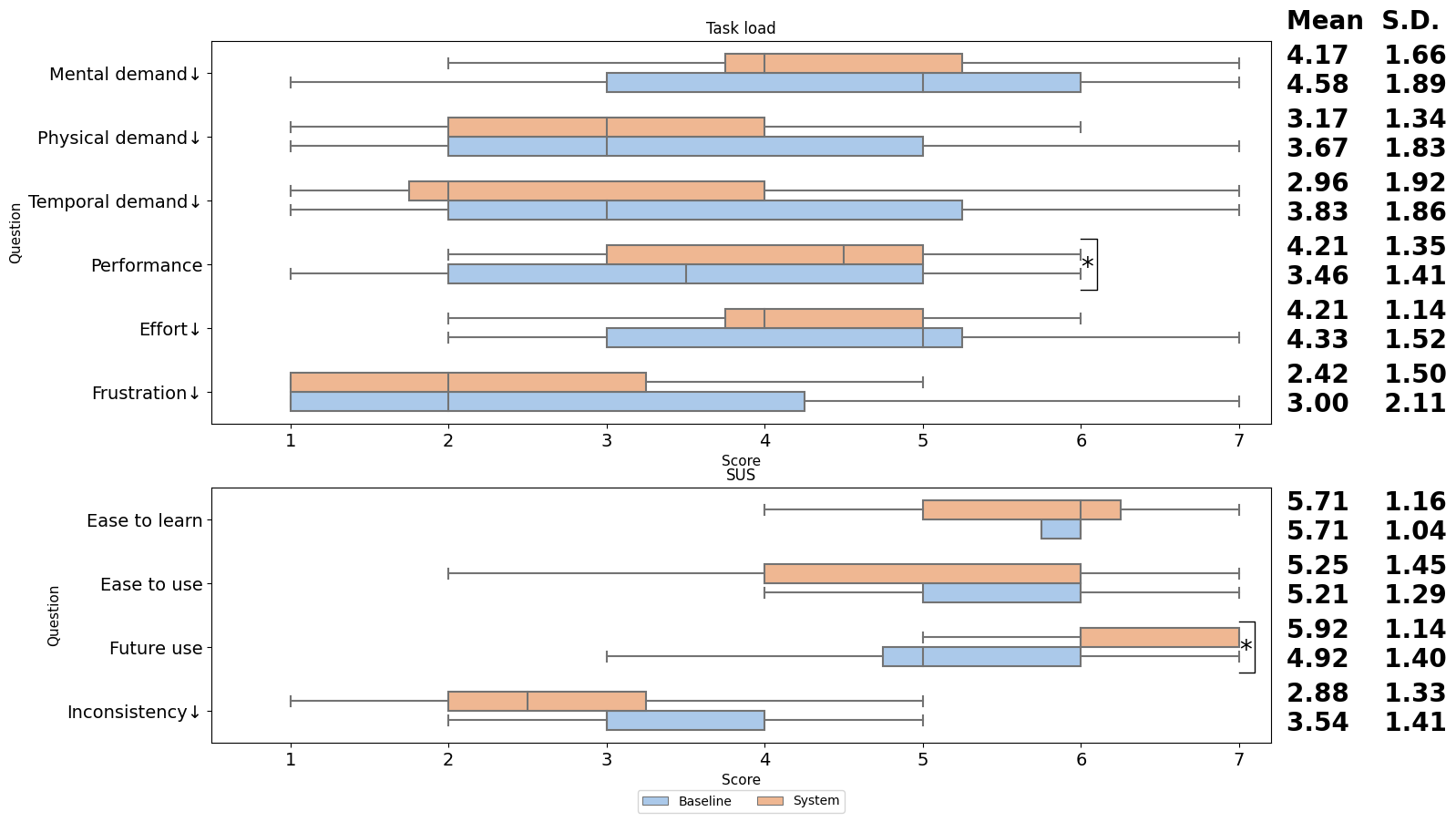}
  \caption{The usability of CoKnowledge and the baseline system. * indicates a significant difference for the item. $\downarrow$  indicates that a lower rating is better. By default, higher ratings are better.}\label{fig:usability}
  \Description{The evaluation of usability of CoKnowledge and the baseline system. The figure shows that despite the additional interactive features, CoKnowledge still maintained comparable usability and task workload levels to the baseline.}
\end{figure*}

Figure \ref{fig:usability} summarizes our assessment of CoKnowledge's usability. %First, we compared the task load between the two conditions. 
Overall, participants reported significantly greater \textbf{success in accomplishing tasks} with CoKnowledge than with the baseline (\textit{Z} = 2.11, \textit{p} = 0.04), while they perceived no statistical difference in other \textbf{task load} items (e.g., mental, physical, temporal demand, and effort). %Although other items did not show significant differences, the mean ratings indicate that the system condition outperformed the baseline condition. 
This suggests that even with the additional features in CoKnowledge, the workload for users was still on par with that of the baseline condition. As discussed in Sections \ref{sec:usefulness} and \ref{sec:interaction}, this is largely achieved by the processing pipeline for danmaku, which considerably reduced the number of comments and effectively structured the collective knowledge.
Regarding \textbf{general usability}, participants were significantly more inclined to use CoKnowledge again (\textit{Z} = 2.29, \textit{p} = 0.02). 
They did not find our system statistically harder to learn or use than the baseline even with the many added features, despite that some users (P16, P22, P23) mentioned that the additional information could be overwhelming.

\section{Discussion}

Previous studies have explored and validated the role of danmaku in constructing collective knowledge \cite{he2021beyond, wu2018danmaku, wu2019danmaku}. 
In this section, we first discuss why our proposed system can improve the utility of danmaku in the digestion of collective knowledge as informed by existing theories. Then, we summarize several design considerations to guide future support for assimilating knowledge co-constructed by media producers and consumers.

% \subsection{The Role of Danmaku \xm{Time-Synced Comments} in Collective Knowledge Assimilation}
\subsection{Strengthening the Role of Time-Synced Comments in Collective Knowledge Assimilation}

% Building on this foundation, our study, informed by existing theories, highlights the significant potential of danmaku to enhance the recall and comprehension of knowledge expressed in both videos and comments \xm{correct?}.

% In our experiment, participants demonstrated improvements in recalling danmaku and video content within the same task duration, which we attribute largely to danmaku itself. 
In our experiment, participants demonstrated improvements in recalling danmaku and video content within the same task duration using our system.
The widely accepted Associative Network Theory (ANT) of memory suggests that human memory functions as a network of bidirectionally linked nodes, where the activation of one node can trigger the retrieval of information in a linked node \cite{collins1975spreading, wheeler2017using, teichert2010exploring, lee2016memory}. The strength of the association between nodes increases the likelihood of successful retrieval \cite{teichert2010exploring}. 
According to this theory, the spatial (overlay on the video) and temporal (synchronized with timestamps) proximity of danmaku and video content turn them into activation nodes for each other. KG provided by CoKnowledge further reinforces these connections. Many participants previously struggled to focus on danmaku, but with CoKnowledge, they could pay closer attention to it, enabling frequent bidirectional activation during the recall phase, even unconsciously \cite{teichert2010exploring}. This helped them recollect video content through linked comments in KG and vice versa. 
P12 and P15, for instance, shared cases where they recalled danmaku first and subsequently retrieved the corresponding video content in the interviews. 
% Given that activation can occur unconsciously \cite{teichert2010exploring}, we hypothesize that most participants subconsciously leverage this bidirectional activation process during recall.

Participants also exhibited significantly better comprehension of collective knowledge with the help of CoKnowledge. 
% , with danmaku playing a crucial role in this process. 
Proficient comprehenders typically excel at processing complex information, identifying key points, and integrating textual structures \cite{mcnamara2009toward, long1993superordinate, singer1992individual}. They are also more adept at making inferences to bridge conceptual gaps and integrating new information with prior knowledge \cite{garnham1982referential, magliano2002using, mcnamara2009toward}. 
% We found that various categories of knowledge danmaku can serve as these comprehension skills. 
We found that \textit{knowledge danmaku} can approximate such comprehension skills, and categorizing them in our system streamlines the process of exercising these skills. 
For instance, \textit{interpretation} danmaku provides concise explanations and summaries of difficult content (PF1, P2, P6).
% , with an example mentioned by P2 shown in Figure \ref{fig:example}). 
\textit{Experience-sharing} danmaku connects scientific concepts to everyday life (PF5). Additionally, \textit{supplementary knowledge} danmaku helps fill knowledge gaps (PF1-4, P2, P8, P10). 
Filtering danmaku comments by category allows users to make up for the skill(s) they lack, significantly increasing the accessibility of scientific content.
% These danmaku comments allow individuals to become skilled comprehenders without barriers, significantly increasing the accessibility of scientific content.

% \begin{figure}[h]
%   \centering
%   \includegraphics[width=0.7\linewidth]{images/example.jpg}
%   \caption{The video clip is explaining the Pauli Exclusion Principle in a rigorous manner, while the highlighted danmaku comment, "In simple terms, neutrons have collision volume, photons don't," offers a more relatable interpretation of the concept.}\label{fig:example}
% \end{figure}

Despite some participants (7/24) mentioning in interviews that they were not accustomed to acquiring knowledge through danmaku, both subjective feedback and objective performance measures indicated notable knowledge gains in the CoKnowledge condition. We argue that participants' discomfort arose from the chaotic presentation of danmaku, which led them to overlook its knowledge-enhancing capabilities. 
We postulate that structural improvements in the presentation and organization of danmaku can practically enhance the exploration of danmaku's knowledge potential.
% Therefore, we advocate for further exploration and utilization of danmaku's knowledge potential.

\subsection{Design Considerations}
% In this subsection, we present several design considerations derived from our analysis to guide future efforts in supporting the assimilation of collective knowledge.
% \xm{Here, you can discuss the generalizability of your findings -- How other types of live comments \xm{e.g., scrolling list in Twitch??} may achieve similar knowledge co-creation between video creators (or live streamers) and viewers \xm{[ref]}. You can then say the below DCs are applicable to general xxx.}
Various forms of live comments can facilitate knowledge co-creation akin to danmaku. For instance, in knowledge-sharing live streams (KSLS), streamers generate real-time content while concurrent viewers actively contribute comments displayed in scrolling lists, dynamically shaping the information flow \cite{lu2018streamwiki, fonseca2021knowledge}. The design considerations generated from the study results, as outlined below, apply to general video communication formats featuring live comments.

\subsubsection{User-Centric Enhancements for AI-Assisted Information Processing}

With the growing volume of UGC online, AI is increasingly used to enhance the efficiency of processing large-scale knowledge \cite{liu2023coargue}. CoKnowledge also relies heavily on AI to process danmaku. However, AI's performance is not flawless; we observed various kinds of errors during our user study. During interviews, some participants reported instances where danmaku were misclassified, leading to confusion (P2) and obstructing comprehension of both the comments and the video content (P12, P17). Additionally, a few participants (P2, P13) noted instances of hallucination in AI-assisted explanations.
% \xm{how about examples of hallucination?} 
To address these issues beyond improving models in the backend, we recommend calibrating user trust and empowering users to co-process information with AI during their interaction with the system. Designers can enhance transparency by visualizing AI uncertainty upon user request \cite{bhatt2021uncertainty} 
% \xm{note that such additional info may increase people's cognitive load.. maybe only display such info when the confidence is below a certain threshold... or when users demand to see them e.g., by hovering over or clicking} 
and help users build a more accurate mental model of the system \cite{gero2020mental}. For example, explaining how the algorithms work \cite{schmidt2020transparency} and highlighting potential limitations \cite{cheng2019explaining} in the system's help information
% \xm{in the system help info? Perhaps not on the main UI...} 
would enable users to leverage the system's knowledge while maintaining critical awareness. Additionally, allowing users to edit AI outputs actively would deepen their understanding of the content through error correction (P3), which, at the same time, can be integrated into knowledge co-construction.
% \xm{correct? any user demanding this feature?}.

\subsubsection{Accommodating Various Strategies While Empowering Users with Full Autonomy}
\label{sec:7.2.2}
McNamara et al. suggested that optimal learning outcomes were achieved with easy information processing and appropriate strategies \cite{mcnamara2009toward}. However, users vary in processing speed, information preferences, and attention allocation, making it difficult to design a "one-size-fits-all" solution. The exceptional performance of CoKnowledge stems not only from reducing the complexity of processing danmaku but also from providing diverse features that accommodate various user strategies. In our experiment, CoKnowledge provided participants with high autonomy, such as the flexibility to switch modes at their own pace and choose whether and how to use specific features. 
Still, our participants demanded more agency. In our current design, CoKnowledge automatically adjusts the ratio between the upper video window and the lower view to
ensure that the part people tend to focus on in the present mode receives more information display space. 
However, a few participants (P11, P22) preferred manual window/view size adjustment to fit their reading needs (discussed in section \ref{sec:use-parttern}). 
% xm{correct? reference the corresponding result section}. 
We thus recommend that knowledge-support tools incorporate a range of adaptable features to address diverse user needs while granting complete control over the strategies users find most comfortable.
% However, one instance where user autonomy was limited—automatic video window resizing when switching modes—drew criticism from a few participants (P11, P22), who preferred manual adjustments. 
% Thus, we recommend that knowledge-support tools incorporate a range of features to address diverse user needs while granting full control over the strategies users find most comfortable.

\subsubsection{Tailoring Feature Designs for Different Knowledge Domains}

The difference between the two video domains in our experiment revealed distinct patterns in how participants utilized the features.
% in face of \xm{xxx knowledge to master}. 
We observed that the AI-assisted explanation feature was rarely used during the health video (0.92 times per person) but was employed more frequently during the astronomy video (5.79 times per person).
% \xm{If you have reported the numbers in the result section, reference; if not, put the numbers here..}. 
This discrepancy can be attributed to the lower complexity of the health video’s danmaku. In contrast, the astronomy video had significantly more \textit{concept-noting} comments, which often triggered the need for additional explanation. Moreover, participants reported using the KG more frequently for the astronomy videos due to their greater content complexity compared to the health video. This trend of feature usage, driven by the nature of the knowledge content, exhibited a predictable and collective pattern. Therefore, instead of building systems that accommodate all knowledge features equally, designers may adopt data structures and representations that best suit a given knowledge domain and prioritize relevant features accordingly. This allows users to efficiently engage with features that maximize the absorption of the specific knowledge source while reducing the learning curve and potential distractions. In resource-constrained scenarios, this approach ensures the most efficient allocation of resources to support knowledge acquisition.

\subsubsection{Navigating the Tension between Information Processing and Natural Co-Viewing Experience}
Evaluation results indicate that CoKnowledge outperformed the baseline in fostering a \textit{"sense of co-presence"} (section \ref{sec:danmaku-util}), primarily through its theme classification and stance analysis. These features allowed users to identify shared perspectives, reinforcing a \textit{"sense of identification"} (P4-6, P12, P24), while also clearly discerning the interplay of contrasting viewpoints, thereby enhancing their \textit{"sense of interaction"} (P4) by simulating dynamic deliberation. However, participants expressed concerns that CoKnowledge's information processing disrupted the \textit{"natural co-viewing experience"} (P20, P23). While the overwhelming volume of danmaku challenges knowledge absorption, it fosters a vibrant atmosphere by conveying the presence of individual co-viewers behind each comment \cite{chen2017watching, liu2016watching}. Casual and colloquial expressions in danmaku, in particular, are a key reason users enable it (PF1, PF5-7, P12, P15). However, CoKnowledge’s knowledge-driven processing pipeline removed much of this entertaining and emotionally resonant content, resulting in a more sterile presentation that diminished participants’ emotional connection to other posters (P20, P23). 
To address this tension, designers should adopt context-sensitive approaches when managing UGC. For users primarily seeking entertainment, retaining unfiltered, original live comments may be preferable. For knowledge-oriented users, CoKnowledge’s processing pipeline is more appropriate. For users with mixed motivations, a hybrid approach may be ideal. For instance, designers could present processed, knowledge-focused content during knowledge-intensive video segments to aid comprehension, while retaining unfiltered comments during less knowledge-intensive segments to preserve the natural co-viewing experience. Alternatively, designers could introduce entertainment-oriented danmaku categories, offering users the option to enable or disable them.

% Participants expressed concerns that CoKnowledge's information processing methods disrupted the natural co-viewing experience (see section \ref{sec:danmaku-util}). While the abundance of danmaku poses many challenges, particularly regarding knowledge absorption, it indeed fosters a vibrant and engaging atmosphere. Emotional and humorous exchanges embedded in danmaku is a key reason many users enable it (PF1, PF5-7, P12, P15). However, CoKnowledge’s knowledge-driven processing pipeline removed much of this entertaining and emotionally resonant content, resulting in a more sterile presentation that diminished participants’ emotional connection to other posters (P20, P23).
% Despite this limitation, CoKnowledge effectively preserved emotional attitudes within danmaku comments through stance analysis, allowing users to clearly perceive contrasting opinions and dynamic interplay of perspectives. This significantly contributed to the superior performance of CoKnowledge in fostering a \textit{"sense of co-presence"}.
% To address such tensions, designers should adopt context-sensitive approaches when managing UGC. If users primarily seek entertainment, it is essential to preserve or enhance emotional and entertaining elements. Additionally, offering multiple modes of information presentation can cater to diverse user needs, allowing users to select modes aligned with their immediate goals—whether for enjoyment or knowledge absorption—and their viewing contexts.

\subsubsection{Alleviating Information Overload}
% Information overload is a prevalent issue that hampers users' ability to effectively engage with danmaku, often prompting some to disable it entirely \cite{ma2017video, wu2019danmaku, wu2018danmaku}.  CoKnowledge effectively alleviated this problem by employing an NLP pipeline to condense danmaku
% and introducing \textit{Focused Mode} that minimized extraneous information. These features allowed users to simultaneously engage with video content and curated high-quality danmaku, facilitating knowledge absorption from both sources (P14, P15, P19).
Information overload hinders users from effectively engaging with danmaku, often prompting some to disable it entirely \cite{ma2017video, wu2019danmaku, wu2018danmaku}. CoKnowledge alleviated this problem by employing an NLP pipeline to condense large volumes of danmaku and enabling users to filter it based on preferences, allowing participants to simultaneously comprehend video content and curated, high-quality danmaku (P14, P15, P19). Additionally, informed by users' viewing habits (Table \ref{formative-study-findings}), it introduced adaptive modes that support manual switching. When users focused on the video itself, focused mode minimized distractions by withholding additional information. Conversely, when users were willing and able to process supplementary information, overview mode and exploration mode provided diverse features, ensuring danmaku remains accessible even when disabled.
Despite these efforts, survey results (section \ref{sec:usability}) showed no significant reduction in task load compared to the baseline. Interviews revealed that participants often felt compelled to engage with features such as Wordstream and KG, even when unnecessary (P2-4, P10, P12-16). These features introduced additional information across different views (P16, P22, P23), thereby increasing cognitive demands for learning to use the system.
This underscores another critical trade-off between information richness and overload. Designers should prioritize maintaining information volume within users' cognitive processing capacities, then selectively implementing features to address user needs without exacerbating task load.

% However, survey results showed no significant reduction in task load compared to the baseline. This can be attributed to CoKnowledge’s additional features designed to support diverse viewing strategies and needs. While these features improved knowledge absorption outcomes (as discussed in section \ref{sec:7.2.2}) and allowed users to process danmaku even when disabled, they inadvertently increased cognitive demands (P16, P22, P23).
% This underscores another critical tension between feature complexity and information overload. Designers should prioritize maintaining information volume within users' cognitive processing capacities. Within this constraint, features should be selectively implemented to address user needs without exacerbating task load.

\subsection{Scalability and Generalizability}
% \xm{[According to 2AC comment 1, first clarify which "results speak only to danmaku systems designed like CoKnowledge" and which parts can be extended]}

% \subsubsection{Applicability and Scalability}
CoKnowledge’s automated pipeline ensures applicability for processing videos of varying durations and danmaku volumes. For newly uploaded videos with insufficient danmaku, CoKnowledge retains its utility by offering progress bar directory (Figure \ref{fig:overview}) and oval nodes in the KG (Figure \ref{fig:exploration}) to organize video knowledge. For highly popular and long-standing videos, the system’s most time-consuming step involves processing substantial pre-existing information, which only requires a one-time effort. This limitation is planned for future optimization. Despite this, CoKnowledge effectively handles the continuous stream of new danmaku these videos attract, leveraging the fine-tuned LLM-based classifier to process them incrementally. Rather than reanalyzing the entire dataset, the system only evaluates the incoming comments to determine whether they qualifies as \textit{knowledge danmaku} and assign them to specific themes. This approach minimizes computational overhead and greatly enhances scalability. Future work could further reduce inference costs by exploring \textit{Programmatic Weak Supervision} (PWS) framework, which has proven effective in domain-specific tasks with limited labeled data \cite{li2024labelaid}.

% \subsubsection{Generalizability beyond Danmaku}
% \label{sec:7.2.1}
Although primarily designed for danmaku-based video platforms, CoKnowledge’s approaches to processing and presenting knowledge are potentially generalizable to platforms featuring traditional comments, such as YouTube \cite{youtube}. These platforms also face challenges from the large volume and unstructured nature of user-generated comments, which can overwhelm users and hamper effective knowledge absorption \cite{liu2022planhelper, liu2023coargue}. CoKnowledge’s processing pipeline—integrating filtering, classification, mapping, and clustering techniques—is well-suited to address these issues by extracting and structuring knowledge from extensive collections of comments.
While CoKnowledge is less tailored to traditional comments addressing overarching video topics, it can effectively present knowledge-related comments targeting specific video scenes. By anchoring these comments to corresponding timestamps through automated alignment methods (e.g., topic modeling, semantic matching) or video referencing interfaces \cite{yarmand2019can, kitayama2008organizing}, designers can seamlessly integrate them into CoKnowledge’s framework, enhancing its adaptability across diverse platforms.

\subsection{Limitations and Future Work}

This study has several limitations. 
% First, our focus on the Bilibili platform may have introduced platform-specific biases. 
First, our data were obtained from the Bilibili platform, which may have introduced platform-specific biases. 
Second, our participants were predominantly young adults (ages 18-31), potentially skewing the findings. Future studies should include more diverse age groups to enhance the accessibility and inclusiveness of CoKnowledge. 
Third, while CoKnowledge filtered out non-knowledge danmaku to increase information density, some participants (P4, P5, P13) reported that eliminating such entertaining danmaku reduced their engagement with the science videos. Future research could explore how non-knowledge danmaku enhance collective knowledge assimilation without compromising content quality.
Besides addressing these limitations, we aim to examine the generalizability of CoKnowledge and its NLP pipeline to other video genres, such as massive open online courses (MOOCs) in the future (P11). 
% \yh{Besides addressing these limitations, future work could expand the analysis to traditional comments, which serve as a complementary channel to danmaku for knowledge sharing \cite{wu2018danmaku, wu2019danmaku}. Integrating these two modalities has the potential to provide a more comprehensive and robust depiction of collective knowledge.}
Moreover, the crowdsourced nature of danmaku offers a self-regulating mechanism that mitigates the spread of misinformation \cite{he2021beyond}. Our design already distinguishes diverse perspectives within danmaku, highlighting its self-correction capabilities. Future research could further explore the effectiveness and patterns of these corrective processes.
Lastly, we plan to conduct a long-term experiment to assess the system's effectiveness over time and explore user strategies within a single video, across multiple viewings, and among different videos.

\section{Conclusion}

In this study, we introduced CoKnowledge, a proof-of-concept tool designed to facilitate the assimilation of collective knowledge in online science videos. Insights from the \hyperref[formative-study]{Formative Study} informed a subsequent \hyperref[content-analysis]{Content Analysis}, which examined the patterns embedded in viewer-generated time-synced comments.
This led to the development of an NLP pipeline for processing danmaku, integrated within an interactive user interface. Results from a within-subject study demonstrated that CoKnowledge significantly enhanced viewers' comprehension and recall of collective knowledge, outperforming an interface displaying unprocessed danmaku in their original forms. We also analyzed usage patterns and gathered feedback on specific design features, guiding the formulation of design recommendations for future tools aimed at supporting knowledge assimilation. Additionally, we encouraged future investigation into the knowledge potential of danmaku.
\section{Acknowledgements}

This work is supported by the Research Grants Council of the Hong Kong Special Administrative Region under General Research Fund (GRF) with Grant No. 16203421. Besides, we would like to thank Haochen Shi, Baixuan Xu, Weiqi Wang for their insightful comments and commitment to the initial design phase. We also extend our gratitude to the reviewers for their constructive feedback, which had significantly improved the quality of our paper.

\bibliographystyle{ACM-Reference-Format}
\bibliography{CHI25}

\end{document}